\documentclass[sn-mathphys-num]{sn-jnl}

\newcommand{\Name}{CTagger}
\usepackage{array}
\usepackage{graphicx}%
\usepackage{varwidth}
\usepackage{tcolorbox}
\usepackage{multirow} 
\usepackage{tabularx}
\usepackage{booktabs}  
\usepackage{array}      
\usepackage{amsmath,amssymb,amsfonts}%
\usepackage{amsthm}%
\usepackage{mathrsfs}%
\usepackage[title]{appendix}%
\usepackage{xcolor}%
\usepackage{textcomp}%
\usepackage{manyfoot}%
\usepackage{algorithm}%
\usepackage{algorithmicx}%
\usepackage{algpseudocode}%
\usepackage{listings}%
\usepackage{subcaption}

\theoremstyle{thmstyleone}%
\theoremstyle{thmstyletwo}%

\theoremstyle{thmstylethree}%

\begin{document}

\title[Article Title]{Identifying Concurrency Bug Reports via Linguistic Patterns}


\author*[1]{\fnm{Shuai} \sur{Shao}}\email{shuai.shao@uconn.edu}

\author[2]{\fnm{Lu} \sur{Xiao}}\email{lxiao6@stevens.edu}

\author[1]{\fnm{Tingting} \sur{Yu}}\email{tingting.yu@uconn.edu}

\affil*[1]{\orgdiv{Department of Computer Science and Engineering}, \orgname{University of Connecticut}, \orgaddress{\city{Storrs}, \postcode{06269}, \state{CT}, \country{USA}}}

\affil[2]{\orgdiv{Department of Systems Engineering}, \orgname{Stevens Institute of Technology}, \orgaddress{\city{Hoboken}, \postcode{07030}, \state{NJ}, \country{USA}}}


\abstract{With the growing ubiquity of multi-core architectures, concurrent systems have become essential but increasingly prone to complex issues such as data races and deadlocks. While modern issue-tracking systems facilitate the reporting of such problems, labeling concurrency-related bug reports remains a labor-intensive and error-prone task.  
This paper presents a linguistic-pattern-based framework for automatically identifying concurrency bug reports. We derive 58 distinct linguistic patterns from 730 manually labeled concurrency bug reports, organized across four levels: word-level (keywords), phrase-level (n-grams), sentence-level (semantic), and bug report-level (contextual). To assess their effectiveness, we evaluate four complementary approaches—matching, learning, prompt-based, and fine-tuning—spanning traditional machine learning, large language models (LLMs), and pre-trained language models (PLMs).  
Our comprehensive evaluation on 12 large-scale open-source projects (10,920 issue reports from GitHub and Jira) demonstrates that fine-tuning PLMs with linguistic-pattern-enriched inputs achieves the best performance, reaching a precision of 91\% on GitHub and 93\% on Jira, and maintaining strong precision on post cut-off data (91\%).  
The contributions of this work include: (1) a comprehensive taxonomy of linguistic patterns for concurrency bugs, (2) a novel fine-tuning strategy that integrates domain-specific linguistic knowledge into PLMs, and (3) a curated, labeled dataset to support reproducible research. Together, these advances provide a foundation for improving the automation, precision, and interpretability of concurrency bug classification.
}


\maketitle

\section{Introduction}
\label{sec:introduction}

Due to the worldwide spread of multi-core architecture, concurrent systems are becoming more pervasive. 
Although concurrent systems are one of the major programming models for improving the performance in the multicore era, 
these systems are prone to concurrency bugs such as data races, deadlocks, and atomicity violations \cite{huang2010leap}. These
software bugs are pervasive, but difficult to deal with due to the complexity of the thread/process synchronization and their non-deterministic nature \cite{bianchi2017reproducing}.
Therefore, different approaches~\cite{yu2016syncprof, yu2018conpredictor, zhang2016lightweight} have been proposed to tackle concurrency bugs.

To track and document software bugs, many modern
software projects use bug-tracking systems (e.g., Bugzilla~\cite{bugzilla}, Google Code Issue Tracker~\cite{gooledoclink}, and Github Issue Tracker
\cite{githublink}). These systems allow developers and users to report issues they have identified in a project. 
Given a bug report, a developer who is assigned to it needs to determine the type of the reported bug, such as whether it is a concurrency bug or not. 
For example, in several projects we have studied, bug reports related to concurrency are explicitly labeled as concurrency bugs.
The labeled bug reports can be useful in many scenarios, such as selecting specific tools to be used for reproducing
the bugs and serving as subjects to be used in the experiments related to concurrency bugs. 
However, manually labeling bug reports is labor-intensive and error-prone (i.e., incorrect label the bug report). 
In addition, to understand the bug, developers often need to look through the bug descriptions, which can be lengthy and verbose. 
 
Therefore, there is a need for an effective approach to reduce the manual effort required to label concurrency bug reports and to identify the cause of the bugs.
The most straightforward way is using keyword search (e.g., "deadlock"), however, research has demonstrated that keyword search tends to be very inaccurate. 
To address this problem, there exist some approaches that use more advanced techniques to automatically classify bug reports or text documents into different categories (e.g., security bugs~\cite{Gegick10, kashiwa2014pilot, ohira2015dataset}, performance bugs~\cite{zhao2020automatically, kashiwa2014pilot}, configuration bugs~\cite{wen2016colua, xia2014automated}, functional bugs~\cite{thung2012automatic, chawla2014automatic}). 
For example, Xia et al.~\cite{xia2014automated} apply n-gram on the description of bug reports to label configuration bug reports related to system setting and compatabilities.
These techniques are based on statistical models (e.g., n-gram). 
There has been some recent work on using deep learning techniques to perform text classification~\cite{kalc,zhao2018investigating, du2020novel, zhang2020text}, such as Convolutional Neural Networks (CNN), Recurrent Neural Networks (RNN), and Transformer model. 
However, existing research does not consider the domain knowledge and may not be able to extract accurate learning features. 

To better understand the domain features to improve the accuracy of bug report labeling, Zhao et al.~\cite{zhao2020automatically} derives a set of linguistic rules related to performance bugs and these rules are used as features to classify performance bug reports. 
However, their approach is targeted at the domain of performance bugs and thus not applicable to classify concurrency bug reports. 
The most relevant work is proposed by Padberg et al.~\cite{padberg2013mining}, who leverage statistical and machine learning techniques to label bug reports related to concurrency bugs. 
They use keywords and the frequencies of keywords as learning features on a linear classifier. 
However, the dataset they used only contains 81 bug reports, with 57 for training and 24 for testing. 
Furthermore, the approach demonstrates a relatively low recall of approximately 47\%, rendering it potentially unsuitable for practical application.

Recent advances in large language models (LLMs) present opportunities for text classification~\cite{wang2023large, zhang2024pushing,sun2023text, wang2024smart} due to their powerful text comprehension capabilities. For example, Wang et al.~\cite{wang2023large} show that LLMs, such as GPT models, can achieve effective zero-shot text classification using prompt strategies, performing competitively with traditional ML and DL methods without labeled training data. Zhang et al.~\cite{zhang2024pushing} propose RGPT, an adaptive boosting framework that iteratively fine-tunes and ensembles base learners, surpassing state-of-the-art models. For bug report classification, Du et al.~\cite{du2024llm} introduce LLM-BRC, utilizing OpenAI’s text-embedding-ada-002 model to classify bug reports in deep learning frameworks with high accuracy (92\%–98.75\%) across TensorFlow, MXNET, and PaddlePaddle. Aracena et al.~\cite{aracena2024applying} explore GPT-based models for automated issue classification, achieving high precision (93.2\%) with minimal training data. 
However, existing works primarily focus on general classification tasks, such as distinguishing bug vs. non-bug reports, rather than addressing domain-specific challenges like concurrency bug classification. While LLMs have shown strong performance in general bug report classification, prompting alone is insufficient for domain-specific tasks. LLMs tend to overgeneralize and infer concurrency where none exists, leading to false positives. Our experiments show that even with structured prompt engineering, LLMs struggle to distinguish true concurrency-related issues from reports that merely mention concurrency-related terms. They often over-infer concurrency involvement, assuming it as the root cause even when it is unrelated to the actual bug. 

In this research, we present the first bug report classification approach specifically designed for concurrency bugs. Unlike general bug classification, domain-specific tasks such as concurrency bug identification require handling unique linguistic characteristics and overcoming dataset imbalance.

Our key insight is that bug reporters tend to use a limited set of domain-specific vocabulary and linguistic patterns when describing concurrency bugs. Based on this observation, we derive a structured set of linguistic patterns (LPs) tailored to concurrency-related issues. LPs capture critical concurrency features across multiple levels—word, phrase, sentence, and context—providing a flexible and effective foundation for classification.
LPs enable diverse classification strategies: they can be directly applied for matching-based classification, used as feature vectors for ML/DL models, leveraged as a guiding tool for LLMs, or integrated into fine-tuned PLMs for enhanced performance. Unlike direct prompting, which relies solely on LLMs’ implicit knowledge, LPs offer explicit and interpretable patterns, ensuring more consistent and accurate classification.
Furthermore, LPs help mitigate dataset imbalance by embedding domain-specific knowledge into learning-based and fine-tuned PLM-based methods.
Our experimental results show that both learning-based and fine-tuned models leveraging LPs significantly outperform direct training without them, achieving higher precision and recall under imbalanced conditions and demonstrating the effectiveness of LPs in real-world concurrency bug classification.

By manually examining 730 concurrency bug reports, we identified a total of 58 linguistic patterns (LPs) across four different levels:
\textbf{Word-level (keywords)}: Using specific keywords to capture individual word-level features, such as terms like “lock,” “thread,” or “race condition,” which frequently appear in concurrency bug descriptions.
\textbf{Phrase-level (n-grams)}: Capturing meaningful patterns by analyzing fixed-length word sequences (e.g., “deadlock occurred” or “multiple threads accessing”).
\textbf{Sentence-level (semantic)}: Identifying semantic patterns that describe concurrency-related actions or issues within a sentence, such as “Thread A acquires the lock but fails to release it,” which implies a resource handling issue.
\textbf{Bug report-level (contextual)}: Analyzing the broader context of the bug report to understand the interplay between multiple sentences or sections, assessing whether the concurrency-related actions are actual causes of the bug.

We next perform a systematic study on utilizing the
linguistic patterns on different classification
models. We categorize them into four categories.
\textbf{Matching-based methods}: The LPs are matched with the target bug report to determine if the report is concurrency-related based on the degree of matching.
\textbf{Learning-based methods}: LPs are treated as feature vectors for various machine learning and deep learning models, enabling the automatic tagging of concurrency bug reports.
\textbf{Prompt-based methods}: LPs are used as prompts to guide large language models (e.g., ChatGPT) in identifying concurrency bug reports.
\textbf{Fine-tuning methods}: LPs are used to fine-tune pre-trained language models to identify concurrency bug reports more effectively. This approach, while more computationally intensive, ensures high precision in identifying relevant reports and offers a thorough analysis with comparable recall. We employ the fine-tuned PLM as \Name{} to automatically identify concurrency bug reports.
To the best of our knowledge, no existing research 
could achieve the same goal. 

\Name{} provides at least three benefits. 
First, developers can label concurrency bug reports in an automated and timely manner. Second, with the 
LPs, \Name{} allows developers to examine the sentences extracted
by matching with the LPs. This can improve the concurrency bug debugging and diagnosis process. Third, researchers can use \Name{} to automatically
search bug reports and thus build datasets for evaluating their techniques and tools.

We conducted a comprehensive evaluation of our approach on 12 open-source projects using two issue-tracking systems—GitHub ($Dataset_{Git}$) and Jira ($Dataset_{Jira}$)—and additionally constructed a post-cutoff dataset ($Dataset_{Post}$) consisting of issues submitted after the LLM’s knowledge cutoff.
In total, the three datasets encompass 10,920 issues, including 546 concurrency-related reports.
\Name{} demonstrated consistently high precision, achieving 91\% on $Dataset_{Git}$, 93\% on $Dataset_{Jira}$, and 91\% on $Dataset_{Post}$.

In summary, this paper makes the following contributions:

\begin{itemize}

\item A set of 58 linguistic patterns spanning four distinct levels for describing concurrency bugs.

\item A systematic study on comparing different classification techniques on utilizing linguistic patterns to identify
concurrency bug reports.

\item The first approach that uses linguistic patterns to fine-tune pre-trained language models (PLMs) for automatically identifying concurrency bug reports.

\item
An extensive empirical evaluation on 10,920 issues from 12 open-source software projects, demonstrating both the effectiveness and generalizability of our approach.

\item
A dataset of labeled bug reports that can be used for replication purposes and future research.

\end{itemize}

In the next section we present a motivating example and background. We then describe the \Name{} approach in Section~\ref{sec:LP}. 
Our empirical study follows in Sections~\ref{sec:evaluation}, followed by discussion in Section~\ref{sec:discussion}. 
We present related work in Section~\ref{sec:related}, and end with conclusions in Section~\ref{sec:conclusion}. 

\section{Motivation and Background}
\label{sec:motivation}

To classify bug reports, 
the most straightforward solution is using key word search (e.g.,
"race", "deadlock"). Almost all existing work on studying concurrency bugs,
when retrieving their corresponding bug reports, use the keyword search
method.  However, our study suggests that a non-trivial
portion of bug reports does not contain any of the commonly
used keywords 
and that some are not concurrency related
bug reports even though they contain the keywords.  In addition,  the performance of these approaches can vary significantly across different projects, depending on the percentage of concurrency bugs present. For example, in our study, we found that the keyword-based approaches achieved a precision of 53.8\% on Hadoop, but only 36.3\% on Spark.

\begin{figure}[H]
\centering
\includegraphics[width=0.85\linewidth]{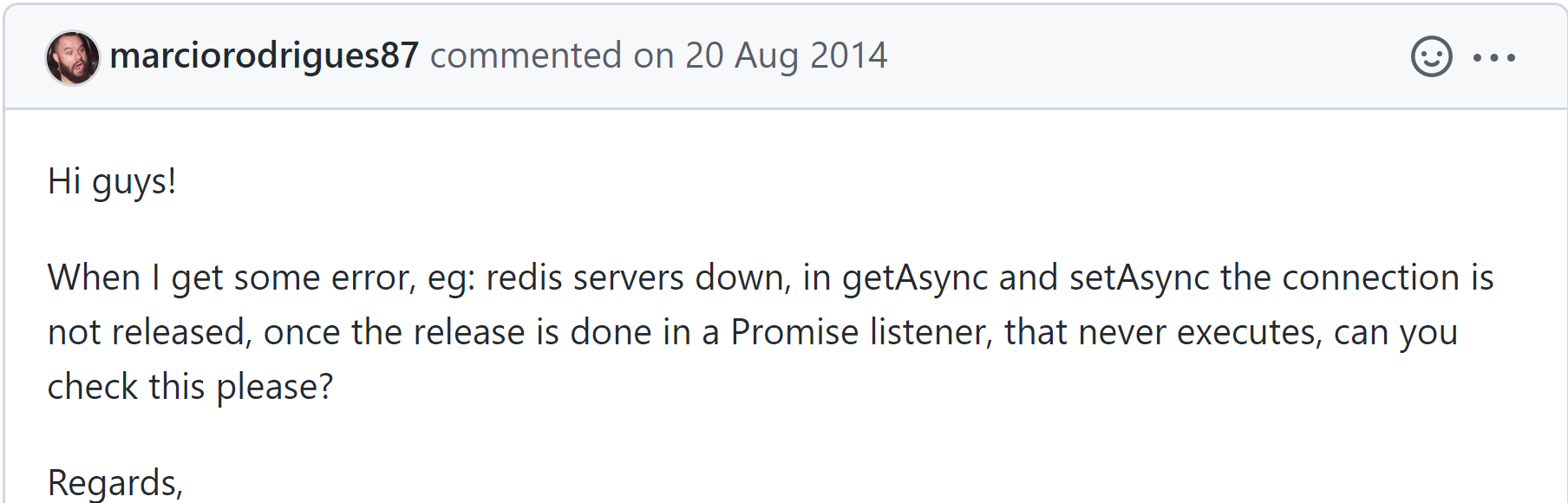}
\caption{False Negative of Keyword Search Method.}
\label{fn}
\end{figure}

\begin{figure}[htbp]
	\centering
	\begin{minipage}{1\linewidth}
		\centering
		\includegraphics[width=0.85\linewidth]{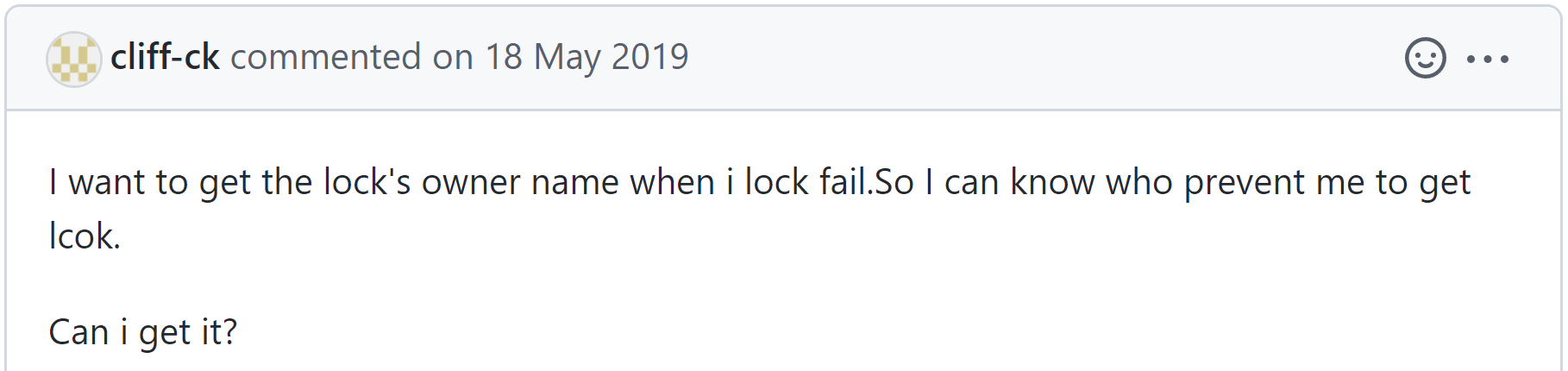}
		\subcaption{Concurrency Related Question.}
		\label{fp1}
	\end{minipage}
	\begin{minipage}{1\linewidth}
		\centering
		\includegraphics[width=0.85\linewidth]{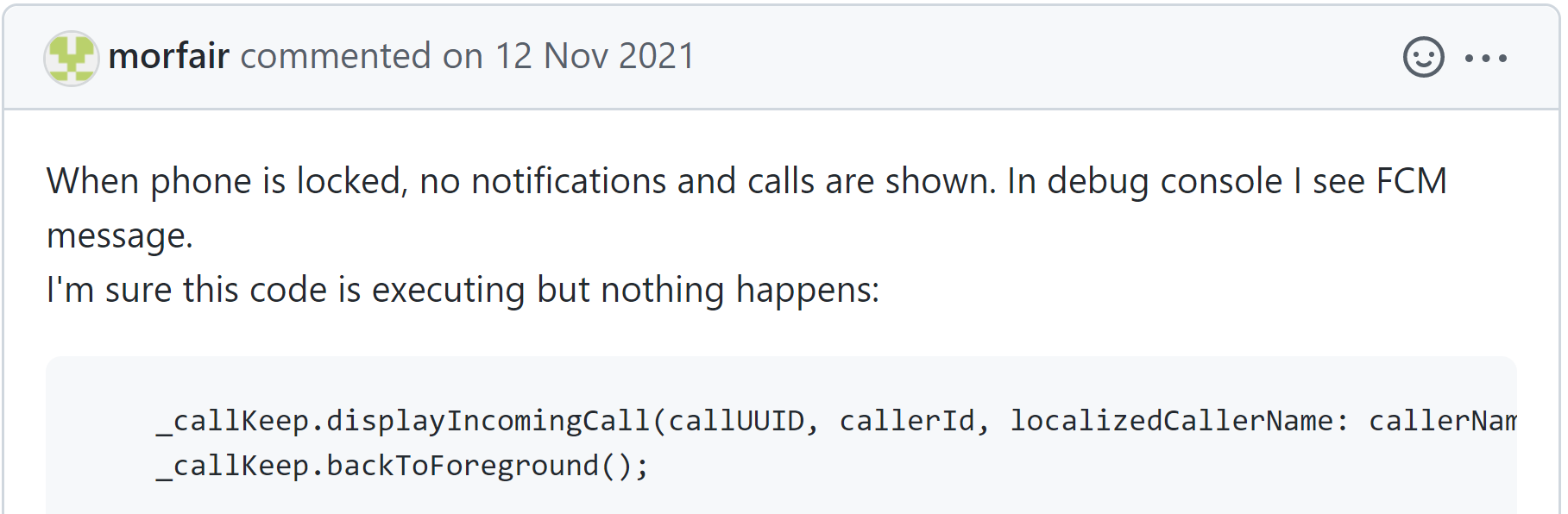}
		\subcaption{Non-concurrency Issue.}
		\label{fp2}
	\end{minipage}
	\caption{False Positive of Keyword Search Method}
	\label{fp}
\end{figure}

Figure~\ref{fn} shows a concurrency bug report. The description of the bug report does not have any common concurrency keywords, but the root cause of this bug
is due to a race condition between the methods \textit{getAsync} and \textit{setAsync}. 
This example shows that simple keywords may miss 
concurrency bug reports (i.e., false negatives).
On the other hand, keyword search can also generate false positive results. When using the common concurrency related keyword ``\textit{lock}" to search 
for concurrency issues issues, Figure \ref{fp1} and 
Figure \ref{fp2} are two matching results. However, Figure \ref{fp1} is  a concurrency-related question
instead of a concurrency bug, although the keyword ``\textit{lock}" appears three times. The reporter asks a question about whether the lock has certain features. 
Figure \ref{fp2} shows a non-concurrency issue, where it describes an Android bug that occurs when the screen of the phone is locked. 

\begin{figure}[H]
\centering
\includegraphics[width=0.85\linewidth]{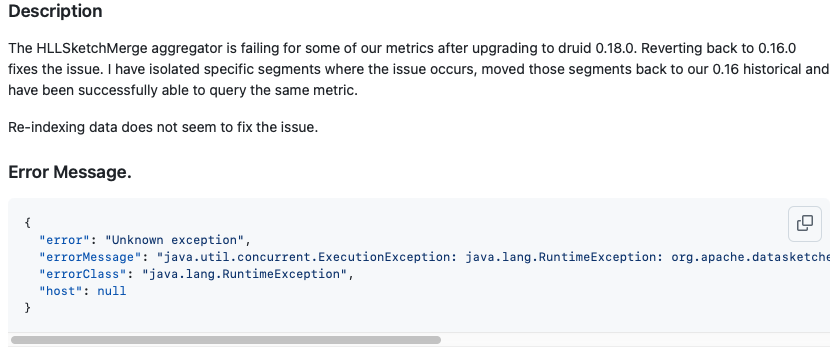}
\caption{False Negative of Prompting GPT-4o.}
\label{fn_g}
\end{figure}

As large language models (LLMs) emerge, models like GPT-4 have demonstrated power reasoning and inference capabilities across various NLP tasks. However, they are not specifically trained for concurrent bug report classification, which introduces significant challenges in identifying concurrency-related issues.  
When explicitly asked to identify concurrency issues, LLMs tend to over-infer concurrency involvement rather than relying solely on factual indicators in the bug report. For example, in Druid-9736 (Figure~\ref{fn_g}), the error message contains `java.util.concurrent.ExecutionException`, which suggests but does not confirm a concurrency issue. GPT-4 incorrectly infers concurrency involvement, sometimes generating non-existent scenarios about threading or synchronization issues, when in reality, this bug was caused by a configuration problem.  

Although LLMs are pre-trained on broad and diverse corpora, they lack exposure to the specialized debugging knowledge required for concurrency-related bug reports.
While they can recognize concurrency-related terms, they often suffer from hallucination—incorrectly inferring concurrency involvement even when it is irrelevant—and thus struggle to distinguish genuine concurrency issues from superficial mentions of terms such as ExecutionException or ThreadPool.
We experimented with various prompting strategies, including Chain-of-Thought (CoT) and Reason-Then-Act (RTA), but even with structured reasoning prompts, LLMs still fail to reliably differentiate true concurrency bugs from non-concurrency issues. Recent work by Koyuncu~\cite{koyuncu2025exploring} empirically evaluated various prompt engineering strategies 
for fine-grained issue categorization (e.g., configuration, network, and security issues) 
and reported that even the best-performing prompting strategy achieved only moderate success 
(F1~$\approx$~0.55 for security-related issues and nearly~0 for network issues). 
These findings highlight the limitations of generic LLM prompting in software issue classification.

The insight of our approach in identifying concurrency
bug reports is that concurrency bugs have 
 unique characteristics ~\cite{torres2018study, abbaspour201710, goetz2006java, artho2003high, ircfl}. Their descriptions 
often contain similar linguistic patterns. 
For example, the occurrence of a concurrency bug 
often involves multiple event entities 
happening in an unintended order.
Therefore, a linguistic pattern capturing
such entities and their ordering 
relationships can be used to identify
this kind of bugs. These patterns can be directly matched, leveraged in ML/DL models, provide valuable insights for prompting LLMs, and be used to fine-tune PLMs for improved classification accuracy.

\subsection{Comparison with Existing Techniques}
\label{backgroud}
Existing work can be classified into four categories. 

\subsubsection{Learning-based approach.}
The first category is using machine/deep learning 
for text classification~\cite{Gegick10, xia2014automated}. Training classifiers
requires labeled datasets, e.g., whether a bug report
is a concurrency bug report or not. 
Gegick et al.~\cite{Gegick10} 
classify bug reports as either security- or non- security-related. 
Xia et al.~\cite{xia2014automated} use text mining to categorize configuration bug reports related to system settings and compatibilities. 
However, none of the above techniques have studied
the characteristics of the concurrency bug reports or propose
solutions to identify them. 

\subsubsection{Pattern-based approach.}
The second approach is using grammar/linguistic patterns to identify 
text belonging to a particular category. The simplest pattern-based
approach is keyword search, in which users provide a list of commonly
used keywords so that any text items (e.g., bug reports) matching with 
the provided keywords will be returned. 

Linguistic patterns have been used in analyzing software engineering natural language 
artifacts to achieve different goals. 
Panichella et al.~\cite{panichella2015can} manually 
identified 246 recurrent linguistic patterns
to classify Android app reviews into 
several categories (e.g., feature request, opinion asking). 
ReCDroid~\cite{zhao2019recdroid} employs linguistic patterns to
extract text related to GUI actions and widgets from bug reports
to support automated bug reproduction. More recently, linguistic patterns have also been applied to smart contract analysis.
SymGPT~\cite{xia2025symgpt} leverages linguistic patterns to extract and formalize ERC rule descriptions from natural language specifications, converting them into symbolic execution constraints for automated detection of rule violations in Ethereum smart contracts. 
There has also been some work on using LPs to analyze bug reports.
For example,
Shi et al. ~\cite{shi2017understanding} presented an approach based on fuzzy method, linguistic patterns, and natural language processing to automatically 
classify the content of feature request issues.
Han et al.~\cite{Han18}  use natural language processing
techniques to 
extract input parameters from textual descriptions of performance bug reports 
and use them to guide performance test generation.

In summary, LPs capture the commonality of a specific 
bug category. Compared with the machine learning approaches,
LPs provide better interpretability and classify texts
without training datasets.

\subsubsection{Combined approach.}
There has been work on combing both LPs and machine/deep 
learning techniques to improve the performance
of classification. 
Chaparro et al. ~\cite{chaparro2017detecting} used linguistic 
patterns to detect missing EB (Expected Behavior) and 
S2R (Steps to Reproduce) in bug descriptions.
Zhao et al.~\cite{zhao2020automatically} derived a set of recurrent linguistic
patterns and combine them with machine learning techniques
to classify bug reports into performance and non-performance
bug reports.  
While existing techniques and approaches
provide different insights and strategies 
to analyze natural language artifacts for
handling software bugs, none of them
have considered using the textual information to address
concurrency bugs.

\subsubsection{LLM-based approach.}
As LLMs emerge, a growing number of studies have explored their applications in text classification. Wang et al.~\cite{wang2023large} demonstrate that LLMs, such as GPT models, can achieve effective zero-shot text classification using prompt strategies, performing competitively with traditional ML and DL methods without requiring labeled training data. Zhang et al.~\cite{zhang2024pushing} propose RGPT, an adaptive boosting framework that enhances LLM-based classification by iteratively fine-tuning and ensembling multiple strong base learners, significantly surpassing state-of-the-art models. Sun et al.~\cite{sun2023text} introduce Clue And Reasoning Prompting (CARP), a method that guides LLMs through progressive clue identification and diagnostic reasoning, improving classification accuracy while addressing token capacity limitations. Wang et al.~\cite{wang2024smart} present a smart expert system that leverages LLMs for text classification, streamlining the process by reducing preprocessing requirements and demonstrating superior performance over traditional methods. 
Several works have also focused on bug report classification. Du et al.~\cite{du2024llm} propose LLM-BRC, which utilizes OpenAI’s text-embedding-ada-002 model to classify bug reports in deep learning frameworks, achieving high accuracy (92\%–98.75\%) across TensorFlow, MXNET, and PaddlePaddle. Aracena et al.~\cite{aracena2024applying} explore GPT-based models for automated issue classification, demonstrating high precision (93.2\%) with minimal training data.

In this work, we first analyze the characteristics of textual descriptions in concurrency bug reports and examine the prevalence of linguistic patterns. Compared to existing studies, we leverage LLMs to generate linguistic patterns more effectively, improving the classification process. Additionally, we adopt a comprehensive approach by integrating linguistic patterns with four different classification methods: matching-based, learning-based, prompt-based, and fine-tuning-based.

\section{Linguistic Patterns for Concurrency Bug Reports}
\label{sec:LP}

The first goal of our approach is to understand how a concurrency bug is reported, which can guide 
us to summarize linguistic patterns (LPs) that reporters 
typically used to describe concurrency bugs. We need to create a dataset with labeled
bug reports, i.e., whether the bug report is related
to a concurrency bug or not. 
However, bug reports with existing labels rarely exist. Therefore, we need to manually 
construct a dataset.
We  describe the process of constructing the dataset
in the following sub-sections.

\subsection{Bug Reports Selection}
\label{selection}

GitHub provides a ``Topics" feature that allows 
users to tag repositories with descriptive words or phrases. 
With topics, users can explore repositories within a particular subject area. 
To ensure that the selected topics comprehensively cover major concurrency-related domains, we referred to prior empirical studies on concurrency bug datasets~\cite{lu2008learning, yu2016syncprof, zhang2016lightweight} and GitHub topic taxonomy. 
Based on this review, we identified seven representative topics frequently associated with concurrency mechanisms—\textit{\#concurrency}, \textit{\#distributed}, \textit{\#high-performance}, \textit{\#mapreduce}, \textit{\#lock}, \textit{\#thread}, and \textit{\#synchronization}. 
By using these tags to search for Java projects, we aim to focus on repositories that align with the subject area of concurrency and offer relevant insights into our study.

We sort the projects within each topic according to their star counts and select the top three repositories per topic. 
Selecting the top three projects balances representativeness and feasibility: repositories with higher star numbers tend to have richer issue discussions and more active maintenance, providing high-quality bug reports while keeping manual annotation tractable. 
Since some projects appear under multiple topics, we obtained a total of 17 unique repositories from this process.

Our goal is to extract English linguistic patterns from as many concurrency bug reports as possible. 
Among the 17 projects, we automatically detected and manually verified the primary language of issue discussions. 
Five projects were excluded because their bug reports were predominantly non-English, which our NLP tools cannot process. 
For the remaining 12 projects, we \emph{manually} checked every issue following the coding process described in Section~\ref{mt}. 
Although some repositories provide coarse-grained labels such as “bug” or “enhancement,” these are inconsistently applied and rarely identify concurrency-specific issues. 
Therefore, manual inspection was necessary to ensure labeling accuracy. 
Four annotators performed the manual review to determine whether an issue was a concurrency bug, as detailed in Section~\ref{mt}.

This multi-topic, multi-project selection strategy ensures that the collected dataset captures diverse concurrency mechanisms—ranging from synchronization and thread management to distributed execution—enhancing the representativeness of the extracted linguistic patterns.

After performing the manual check, the number of issues 
in the 12 projects ranged from 37 to 6,133. 
However, in five out of the 12 projects, 
none of the issues were related to concurrency bugs. 
This observation suggests that many repositories tagged with concurrency-related topics may focus on concurrency-safe frameworks or configurations, rather than containing actual concurrency defect reports. 
As a result, seven projects remained in our study: 
\textit{Druid}, \textit{gRPC-java}, \textit{Presto}, \textit{Pulsar}, 
\textit{Redisson}, \textit{Trino}, and \textit{Vert.x}. 
These projects cover a diverse range of domains and offer valuable contributions to the field of software development. 
\textit{Druid}~\cite{druid} stands out as a powerful data store optimized for real-time analytics, specifically for handling large-scale, time-series data. 
\textit{gRPC-java}~\cite{grpc-java}, on the other hand, provides a high-performance framework for building distributed systems and 
microservices using RPC technology. 
\textit{Presto}~\cite{presto} and \textit{Trino}~\cite{trino}, formerly known as PrestoSQL, excel in distributed SQL query processing, 
enabling fast and interactive analytics on big data. 
\textit{Pulsar}~\cite{pulsar} offers scalable messaging and event streaming capabilities, 
facilitating real-time data processing and event-driven architectures. 
\textit{Redisson} serves as a Java library that simplifies the utilization of Redis for distributed caching and data manipulation tasks. 
Lastly, \textit{Vert.x}~\cite{vert.x} emerges as a lightweight and event-driven toolkit for building reactive applications, 
empowering developers to create scalable and concurrent applications across various domains. 
Notably, these seven projects are also used in existing studies on concurrent fault localization and benchmarking~\cite{ircfl}, highlighting their relevance and representativeness for concurrency research. 
Together, these projects span multiple concurrency paradigms—including distributed processing, reactive programming, and synchronization-intensive architectures—providing a diverse and representative foundation for linguistic pattern extraction.

In this work, we have adopted a strategic approach to select concurrency bug reports from specific projects rather than extracting them from the entirety of GitHub. 
Utilizing a keyword-based search across all GitHub repositories, while seemingly exhaustive, bears intrinsic limitations. 
Prior studies on automated issue retrieval have also reported that keyword-based search often suffers from both low recall and semantic mismatch, especially for domain-specific bugs such as concurrency issues~\cite{xia2014automated, zhao2020automatically}. 
Notably, not all concurrency bugs are explicitly labeled with precise keywords, and there is a substantial risk of omitting relevant reports. 
Furthermore, the choice of keywords is pivotal and introduces bias, as it predicates the search on predefined assumptions about the nature of concurrency bugs.  
In contrast, focusing on well-maintained projects allows the collection of high-quality, context-rich issue discussions that provide clearer descriptions of concurrency symptoms and root causes.  
By concentrating our efforts on specific projects, we aim to foster a dataset that encapsulates a rich array of concurrency bug reports, enhancing the accuracy and reliability of our subsequent analyses and findings.

Additionally, searching for bug reports directly from various bug tracking systems can be a time-consuming and labor-intensive process. 
With numerous projects and a vast array of bug reports to consider, manually checking each report would impose a significant workload on the researchers. 
To ensure the feasibility and efficiency of our study, we opted to focus on popular and widely-used projects that are known to involve concurrent programming. 
This targeted approach allows us to concentrate our manual checking efforts on a manageable yet representative set of bug reports, increasing the likelihood of identifying relevant concurrency issues.

In this paper, we focus on projects written in Java. 
We acknowledge that different programming languages have their own concurrency-related characteristics. 
However, the goal of our study is to demonstrate the effectiveness of our approach within the context of Java projects. 
By concentrating on a specific programming language, we were able to delve deeper into language-specific nuances and exploit Java-specific concurrency features to enhance our bug report classification technique. 
Focusing on Java also ensures consistency in concurrency semantics (e.g., thread model and synchronization primitives), which facilitates reproducibility and comparability across projects.  

It is possible to extend our approach to projects written in other programming languages to broaden its applicability. 
However, such an extension would require adapting our linguistic patterns to account for language-specific concurrency mechanisms.  
For example, in Java, concurrency bugs are often related to threads and synchronization; 
in Python, they often involve the Global Interpreter Lock (GIL); 
in C++, they commonly arise from low-level primitives such as mutexes and semaphores. 
Therefore, the description and manifestation of concurrency bugs vary across programming languages.  
Nonetheless, our approach is language-agnostic in design and can be generalized to other ecosystems with appropriate adaptation.

\begin{table}[]
    \caption{Selected projects}\label{projects}
    \centering
    \begin{tabular}{|c|c|c|c|}
    \hline
    \textbf{Projects}  & \textbf{\#Issues} & \textbf{\#ConBugs}  & \textbf{\#ConSents} \\ \hline
    Druid     & 4540  & 127        & 428                \\ \hline
    gRPC-Java & 3098  & 96         & 289                \\ \hline
    Presto    & 5283  & 60         & 191                \\ \hline
    Pulsar    & 6133  & 156        & 466                \\ \hline
    Redisson  & 4378  & 279        & 629                \\ \hline
    Trino     & 5022  & 62         & 166                \\ \hline
    Vert.x    & 2528  & 132        & 384                \\ \hline
    \textbf{Total} & 30982  & 912  & 2553               \\ \hline
\end{tabular}
\end{table}

Table 1 shows the statistics for each project,
including the number of total issues (\#Issues), 
the number of concurrency bug reports (\#ConBugs), 
the number of concurrency related sentences (\#ConSents).
The data was collected up to the date March 14, 2022.

\subsection{Manual Tagging}
\label{mt}

We manually tag all bug reports from the collected 
bug reports as either concurrency bugs or non-concurrency bugs. We also tag every sentence in each concurrency bug report as concurrency-related or not concurrency-related. 
The reason for sentence-level tagging is that
not all sentences in a concurrency bug report are related
to the bug, such as those describing the environment
setup. Therefore, these sentences may not help with the identification of concurrency bug reports.
In contrast, we tag all bug-related sentences,
which describe the step-to-reproduce, the symptom,
and the expected behavior
of the bug.
The linguistic patterns are also extracted from the bug-related sentences.
Because linguistic patterns are applicable to natural language (NL), 
we only consider the NL part of the bug report,
including the bug report title and the NL content of the original post.

Specifically, to label bug-related sentences, we consider the Observed Behavior (OB), the Steps to Reproduce (S2R), and the Expected Behavior (EB), because the three parts are essential to the 
completeness of a bug report~\cite{chaparro2017detecting}. 
The OB would describe the errors, exceptions, warnings, and problems being observed caused by the bug. 
The EB in a concurrency bug report would describe the expected behavior of the correct program. 
The S2R describes the steps to reproduce the reported bug. 
The sentences of the bug report can belong to any of the three parts. 

\subsubsection{Coding process.}
Four annotators were involved in the labeling of bug reports and sentences, 
consisting of two graduate students and two undergraduate students. 
All annotators had at least four years of software development experience, 
including practical exposure to concurrent system development and debugging.

To ensure the reliability of the coding process, a pilot study was conducted by one of the graduate students on 200 randomly selected bug reports. 
The purpose of the pilot study was to establish clear and consistent coding guidelines for accurately classifying bug reports. 
The coding guideline defined concrete linguistic and contextual indicators for identifying concurrency bugs, such as mentions of synchronization primitives (e.g., “lock,” “synchronized”), thread interactions (“race,” “thread A/B”), and concurrency-related symptoms (e.g., “deadlock,” “hang,” “interleaving”).  
It also included negative examples (e.g., “UI thread,” “lock screen”) that should not be labeled as concurrency-related, ensuring consistent boundary judgments across annotators.  
The outcome of the pilot study included these guidelines, annotated examples, and training materials, which were later used to instruct all annotators.

Subsequently, the lead annotator conducted a one-hour training session for the other three annotators, introducing the pilot results and discussing the detailed labeling procedures. 
This training session aimed to ensure a shared understanding of the guidelines and a uniform interpretation of concurrency-related cues.  

To distribute the 30,982 bug reports evenly, we formed two annotation groups, each consisting of two annotators.  
Bug reports were evenly and randomly assigned to the groups, giving each annotator responsibility for 15,491 reports.  
Within each group, annotators independently labeled the same set of issues and then reconciled differences through discussion to reach consensus.  

During the initial independent labeling phase, the average pairwise agreement among annotators was approximately 68.4\%, which improved to 72.8\% after intra-group discussion, demonstrating the benefit of consensus-based refinement.  
Once both groups completed labeling their respective subsets, a cross-validation phase was conducted: each group reviewed the labels assigned by the other group to their set of 15,491 reports.  
This cross-validation step served as an inter-group consistency check, ensuring that labeling standards were uniformly applied across the entire dataset.  
Most disagreements arose from cases where concurrency was indirectly implied (e.g., thread-pool misconfiguration or asynchronous callbacks).  
Conflicts were resolved in weekly 30-minute meetings, where annotators reviewed representative examples and reached consensus under the supervision of a senior author.  
A bug report was ultimately labeled as concurrency-related only if all four annotators unanimously agreed after cross-validation and discussion, ensuring high labeling precision and reproducibility.  

\textbf{Cohen’s Kappa coefficients.}  
The entire annotation process required approximately 640 person-hours, including both issue-level and sentence-level tagging.  
We used Cohen’s Kappa coefficient to measure inter-group agreement.  
The coefficient for issue-level tagging was 72.8\%, and for sentence-level tagging 85.7\%, both indicating substantial agreement~\cite{mchugh2012interrater}.

\subsection{Data Pre-processing}
\label{dapre}

We split each bug report into sentences
because the concurrency linguistic patterns (LPs)
are generated at the sentence level.
We analyze the words and the
grammatical structure of each sentence to capture information relevant to LPs. 
This is because certain words frequently co-occur within a single sentence. 
For example, 
in the sentences ``\textit{... the lock hangs forever}" and ``\textit{... thread is hung}," 
the verb \textit{hang} often appears with concurrency-related nouns such as \textit{lock} or \textit{thread}. 
Therefore, we pre-process the
bug report datasets by splitting them into sentences
and categorizing their words by linguistic roles (e.g., nouns, verbs). 

We leverage SpaCy~\cite{SpaCy} to segment each report into sentences and use the manually labeled sentence-level annotations described in Section~\ref{mt}. 
For each concurrency-related sentence, we apply several natural language processing (NLP) techniques. 
First, we perform lemmatization and Part-Of-Speech (POS) tagging on each sentence to obtain the base form and POS tag for every token.  
Because these sentences are concurrency-related, the contained words often encode concurrency semantics. 
We group words according to their POS category, forming sets such as \textit{VERB} and \textit{NOUN}.  

Unlike regular text, bug reports frequently contain code-related entities (e.g., classes, methods, and interfaces), which also play an essential role in identifying concurrency bugs. 
For example, in the sentence ``\textit{tryLock() throws Exception!}," \textit{tryLock()} is a concurrency-related API and its failure suggests a potential concurrency bug. 
However, code entities are difficult to recognize using standard POS or NER techniques.  
Prior studies have shown that general-purpose NER models often fail to capture programming entities in mixed natural-language/code text~\cite{tabassum2020code}, motivating the use of software-specific NER in our setting.  
To this end, we use a software-specific NER~\cite{tabassum2020code} model trained on 152 million sentences from StackOverflow~\cite{gousios2012ghtorrent} to recognize and classify classes, packages, methods, interfaces, and other code artifacts in each sentence. 
We then save these recognized entities into the set \textit{API}.  

To ensure preprocessing reliability, we randomly sampled 200 pre-processed sentences to manually verify SpaCy’s segmentation and NER recognition results, achieving over 95\% correctness in both tasks.  
The resulting pre-processed data is represented as a tuple $\langle$Sentence, VERB, NOUN, API$\rangle$, which serves as the foundation for linguistic pattern extraction in Section~\ref{sec:LP}.

\subsection{Deriving Concurrency Linguistic Patterns}

Among all 912 labeled concurrency bug reports (Table \ref{projects}), 
we used 80\% (730 reports) containing 1,734 concurrency-related sentences to extract linguistic patterns (LPs), 
and reserved the remaining 20\% (182 reports, 817 sentences) to evaluate the effectiveness of the extracted patterns.

Drawing inspiration from prior work~\cite{zhao2020automatically, shi2017understanding}, 
we define four distinct levels of linguistic patterns (LPs) to characterize concurrency bug reports: 
word-level (keywords), phrase-level (n-grams), sentence-level (semantic), and bug-report-level (contextual). 
We adopted a similar foundational protocol across these levels, while accounting for the unique characteristics of concurrency-related bug reports. 
For instance, as shown in~\cite{zhao2020automatically}, profiling patterns designed for performance bugs are not directly applicable to concurrency-related issues.

\begin{table}[]
\caption{Different Types of Words }\label{definition}
\centering
\begin{tabular}{|l|l|l|l|c|}
\hline
\textbf{Abbr} & \textbf{Types}                         & \textbf{Words}                                     & \textbf{POS}  & \textbf{\#words} \\ \hline
CBG  & Concurrency Bug               & deadlock, livelock, race condition, etc.  & NOUN & 10 \\ \hline
CME  & Concurrency Mechanism         & lock, thread, writelock, transaction, etc.     & NOUN & 6 \\ \hline
CTR  & Concurrency Terminology       & concurrency, synchronization, etc.        & NOUN\&ADJ & 29 \\ \hline
POP  & Programming Operation         & lock, fork, hold, require, release, etc.  & VERB & 43 \\ \hline
PSY  & Programming Symptom           & hang, stuck, block, remain, freeze, etc.  & VERB & 34 \\ \hline
AOT  & Adverb of Time                & again, forever, already, infinitely, etc. & ADV\&ADJ & 23 \\ \hline
SYB  & Synonym for Bug               & bug, fail, issue, problem, error, etc.        & NOUN\&ADJ & 26 \\ \hline
API  & Specific-API                  & lock(), unlock(), Thread.sleep(), etc.        & API  & 32 \\ \hline
EXC  & Exceptions                    & NullPointerException, etc.  & API  & 11\\ \hline
NEG  & Negative Words                & no, not, never, incorrectly, etc              & ADV\&ADJ  & 21\\ \hline
\end{tabular}
\end{table}

\subsubsection{Manual Coding}

To extract linguistic patterns across four levels, we followed a systematic procedure combining manual analysis with GPT-4o, 
the state-of-the-art large language model available from OpenAI at the time of this study, to ensure both accuracy and scalability. 
This process began with identifying concurrency-related sentences and progressed through iterative refinements to categorize patterns into different levels. 
To ensure consistency and mitigate hallucination, we used a fixed set of structured prompts across all GPT-4o runs, 
which included explicit instructions to verify each candidate pattern against the original sentence context. 
Annotators manually reviewed and approved all GPT-generated outputs before inclusion.
The complete set of prompts is provided in our replication package~\cite{shao2025replication}.

\noindent
\textit{Step 1: Identifying Concurrency-related Sentences.}  
Following the procedure described in Section~\ref{mt}, two groups of annotators analyzed 912 labeled concurrency bug reports.  
Each group was assigned 456 reports to tag and extract linguistic patterns.  
Annotators reviewed each bug report to identify sentences specific to concurrency bugs, which served as the foundation for subsequent pattern extraction.

\noindent
\textit{Step 2: Extracting Word-level Patterns (Keywords).}  
Annotators systematically examined each sentence in the corpus, tokenizing them into individual words and performing Part-of-Speech (POS) tagging to determine grammatical categories (e.g., verbs, nouns).  
Words semantically related to concurrency were collected along with their POS tags.  
This step yielded a set of concurrency-related keywords, forming 23 word-level patterns as summarized in Table~\ref{patterns}.

\noindent
\textit{Step 3: Extracting Phrase-level Patterns (N-grams).}  
Building upon the word-level results, annotators analyzed co-occurrence frequencies to identify recurring multi-word sequences (n-grams).  
By examining how concurrency-related words frequently appear together, we derived phrase-level patterns that capture meaningful relationships and typical concurrency scenarios.  
Each candidate pattern was required to appear in at least 5\% of the corpus and to be independently confirmed by two annotators for inclusion.

\noindent
\textit{Step 4: Extracting Sentence-level Patterns (Semantic).}  
To extract sentence-level patterns, we leveraged GPT-4o for semantic analysis.  
GPT-4o received concurrency-related keywords and phrases from earlier steps and analyzed the sentences containing them to determine whether they genuinely described concurrency phenomena.  
Annotators then manually verified and refined GPT-4o’s outputs, summarizing 17 sentence-level patterns that represent how concurrency bugs are described in natural language (Table~\ref{patterns}).

\noindent
\textit{Step 5: Extracting Bug Report-level Patterns (Contextual).}  
At the bug report level, our goal was to determine whether the concurrency-related sentences directly corresponded to the bug’s root cause.  
GPT-4o analyzed the contextual relationships between concurrency descriptions and the overall bug report narrative, while annotators validated and refined its outputs to ensure that the resulting patterns accurately captured causal relationships.

During manual verification, annotators achieved an average agreement rate of 82\% on the inclusion of candidate patterns, indicating consistent understanding of concurrency-related linguistic expressions.  
In total, we derived 58 linguistic patterns across four levels: 23 word-level, 12 phrase-level, 17 sentence-level, and 6 bug report-level patterns.

\subsubsection{Word-level LPs}

In our study, we categorized keywords into ten distinct sets to comprehensively capture the diverse terminologies appearing in bug reports.  
The initial categorization was informed by prior research on bug report classification~\cite{xia2014automated, zhao2020automatically} and further refined through manual inspection of representative reports from our dataset.  
These ten sets serve as the foundation for constructing higher-level linguistic patterns.  
For the word-level LPs, our primary goal is to capture the core linguistic characteristics that distinguish concurrency bugs from other issue types.  
Accordingly, we focus on two key categories—\textit{Concurrency Bug} (CBG) and \textit{Concurrency Mechanism} (CME)—which encompass terms directly associated with concurrency phenomena such as \textit{deadlock}, \textit{race}, and \textit{lock}.  
Together, these two categories comprise 23 keywords and short phrases explicitly describing concurrency issues, making them essential for the targeted, keyword-based identification of concurrency bug reports.

To extract word-level linguistic patterns, we employed a frequency-based threshold to filter out patterns with low occurrence rates in the dataset. 
The threshold was varied between 0\% and 20\% to analyze its impact. 
To balance coverage and precision, we empirically evaluated the classification performance across different thresholds. 
After careful analysis, we determined that setting the threshold at 5\% (i.e., retaining patterns that appear in at least 5\% of concurrency-related sentences) yields the best trade-off. 
This choice allows us to preserve patterns that are both statistically significant and representative of concurrency-related bug descriptions while filtering out infrequent, project-specific noise. 

\subsubsection{Phrase-level LPs}

A phrase-level linguistic pattern (LP) extends basic keyword matching by incorporating additional bigram and trigram phrases ($n \in \{2,3\}$). 
We limited our analysis to bigrams and trigrams because higher-order n-grams often resulted in sparse or overly specific phrases that did not generalize across projects, 
while $n=2$ or $3$ provided the best balance between semantic richness and statistical reliability.

In our study, we observed that certain words and phrases frequently co-occur in concurrency-related sentences. 
Since the concurrency-related words listed in Table~\ref{definition} are categorized into four specific POS tags—NOUN, VERB, ADV, and API—we extracted phrases associated with these tags using the POS pattern mining technique implemented in SpaCy~\cite{SpaCy}. 
The POS pattern mining module identifies consecutive token sequences that match predefined syntactic structures such as NOUN–NOUN, ADJ–NOUN, or VERB–NOUN, 
which often express action–object or compound relationships common in concurrency contexts (e.g., “acquire lock”, “thread pool”).

After extracting candidate bigram and trigram phrases, we associated each phrase with its corresponding category type (as defined in Table~\ref{patterns}). 
For example, in the sentence “\textit{Thread deadlock in stress test!},” six bigram phrases were extracted, including \textit{thread deadlock}, \textit{thread stress}, \textit{thread test}, \textit{deadlock stress}, \textit{deadlock test}, and \textit{stress test}. 
Among these, only \textit{thread deadlock} was retained, since both \textit{thread} and \textit{deadlock} are NOUNs belonging to the CME and CBG categories respectively, 
yielding the PH1 pattern illustrated in Table~\ref{patterns}.

To ensure relevance, we retained only phrases containing at least one concurrency-related term (CBG or CME) and removed non-concurrency expressions (e.g., “stress test”). 
Annotators manually reviewed the filtered set to confirm phrase validity. 
The final phrase-level LPs comprised 12 recurring patterns that appeared in more than 10\% of concurrency-related sentences, 
providing a richer syntactic representation of concurrency-related descriptions compared to isolated keywords and enhancing the precision of pattern-based analysis.

\subsubsection{Sentence-level LPs}

Sentence-level linguistic patterns (LPs) aim to capture the semantic meaning of sentences containing predefined concurrency-related words or phrases.  
As noted earlier, sentences mentioning concurrency terms do not always convey concurrency semantics (e.g., a sentence containing “lock” may not describe synchronization behavior).  
Traditional dependency parsers can extract grammatical roles (e.g., subject–verb–object) but cannot infer the underlying concurrency intent—such as distinguishing “lock is held forever” from “lock is required here.”  

To address this limitation, we leveraged large language models (LLMs) for semantic interpretation, using their contextual reasoning capabilities to complement syntactic analysis.  
Among available LLMs, we selected \texttt{GPT-4o} as the primary implementation due to its strong performance on semantic understanding tasks and its favorable cost–efficiency balance compared with other models (e.g., Claude and GPT-4).  
The model was used to infer concurrency semantics—such as lock acquisition, thread blocking, or race symptoms—within sentences containing concurrency-related expressions.  
For example, in the sentence “Thread A acquires the lock but fails to release it,” the LLM recognizes the implicit concurrency action pattern.

To mitigate potential hallucination, we adopted a three-layer control strategy:  
(1) constrained prompts restricted the LLM’s output to a fixed set of concurrency-related categories (e.g., lock action, lock symptom, thread operation);  
(2) dual independent validation ensured consistency across annotators; and  
(3) inconsistent or unsupported generations were discarded.  
Through iterative refinement—where the LLM proposed candidate semantics and annotators consolidated or pruned them—we derived 17 stable sentence-level patterns that capture recurrent concurrency-related semantics.

In practice, these finalized patterns serve as structured prompts to guide LLM-based classification.  
This design transforms the LLM from an open-ended generator into a controlled semantic classifier, leveraging its contextual understanding while maintaining reproducibility and minimizing hallucination.

\subsubsection{Bug Report-Level LPs}

Bug report-level linguistic patterns (LPs) capture the contextual role of concurrency-related sentences within an entire bug report.  
While a sentence may describe a concurrency phenomenon (e.g., “acquires a lock”), it does not necessarily correspond to the bug’s root cause.  
Thus, we frame the task as a contextual reasoning problem—determining whether a concurrency-related sentence directly contributes to the reported failure.

Directly classifying bug reports into specific root-cause categories (e.g., race condition, deadlock) often produces unreliable results because large language models (LLMs) lack specialized debugging experience.  
To mitigate this limitation, we reformulate the problem as a general contextual judgment task, asking the LLM whether a given concurrency-related sentence represents the underlying cause of the bug.  
Among available models, we employ \texttt{GPT-4o} for its balance of reasoning capability and cost efficiency.

In this process, the LLM receives concurrency-related sentences alongside their full bug report context and infers whether each sentence describes a root cause.  
Annotators then validate these outputs and consolidate consistent cases into higher-level bug report patterns—such as lock-related or thread-related issues.  
Fine-grained categories (e.g., lock action vs.\ lock symptom) are merged when multiple concurrency sentences collectively describe the same underlying fault.  
For example, reports containing both lock-acquisition failures and deadlock symptoms are grouped under “lock-related issues.”

To prevent hallucination, we constrain the LLM’s outputs to a predefined set of contextual categories and require independent human verification.  
This hybrid design combines the semantic reasoning capability of LLMs with human oversight, producing interpretable, domain-grounded, and reliable bug report–level linguistic patterns.

For a comprehensive overview of the prompts used in our study and the complete set of linguistic patterns identified across all levels, please refer to our replication package~\cite{shao2025replication}.

\begin{table}[h!]
\centering
\caption{Summary of Linguistic Patterns}
\label{patterns}
\begin{tabularx}{\textwidth}{l|l|l|X}
\hline
\textbf{Level} & \textbf{ID} & \textbf{Linguistic Patterns} & \textbf{Example\textbackslash Description} \\ \hline

Word & KW & CBG$|$CME & Basic keyword match \\ \hline

\multirow{6}{*}{Phrase} 
& PH1 & CBG$_{[NOUN]}$+CME$_{[NOUN]}$ & \multirow{6}{*}{\parbox{6cm}{Basic keyword match concept by additionally considering trigram/bigram phrases.}} \\
& PH2 & CBG$_{[NOUN]}$+SYB$_{[NOUN]}$ & \\ 
& PH3 & CME$_{[NOUN]}$+POP$_{[VERB]}$ & \\ 
& PH4 & CME$_{[NOUN]}$+PSY$_{[VERB]}$ & \\ 
& PH5 & CME$_{[NOUN]}$+SYB$_{[NOUN]}$ & \\ 
& \dots & \dots & \\ \hline

\multirow{3}{*}{Sentence}
& SE1 & Lock Action & I am trying to acquire a fair lock. \\ 
& SE2 & Lock Symptom & The system hangs when it tries to acquire a lock. \\ 
& \dots & \dots & \dots \\ \hline

\multirow{3}{*}{Bug Report}
& BR1 & Lock Issue & The root cause is lock action/lock symptom. \\ 
& BR2 & Thread Issue & The root cause is thread action/thread symptom. \\ 
& \dots & \dots & \dots \\ \hline

\end{tabularx}
\end{table}

\subsection{Identifying Concurrency Bug Reports}

\label{sec:IV}

We examine LP-based methods across four categories: matching-based methods, learning-based methods, prompt-based methods, and fine-tuning-based methods. After a comprehensive evaluation, the fine-tuning-based approach demonstrates the highest performance (see Section 5). Therefore, we adopt the fine-tuned model as \Name{}, a default solution for identifying concurrency bug reports.

\begin{figure*}
\centering
\includegraphics[width=1\textwidth]{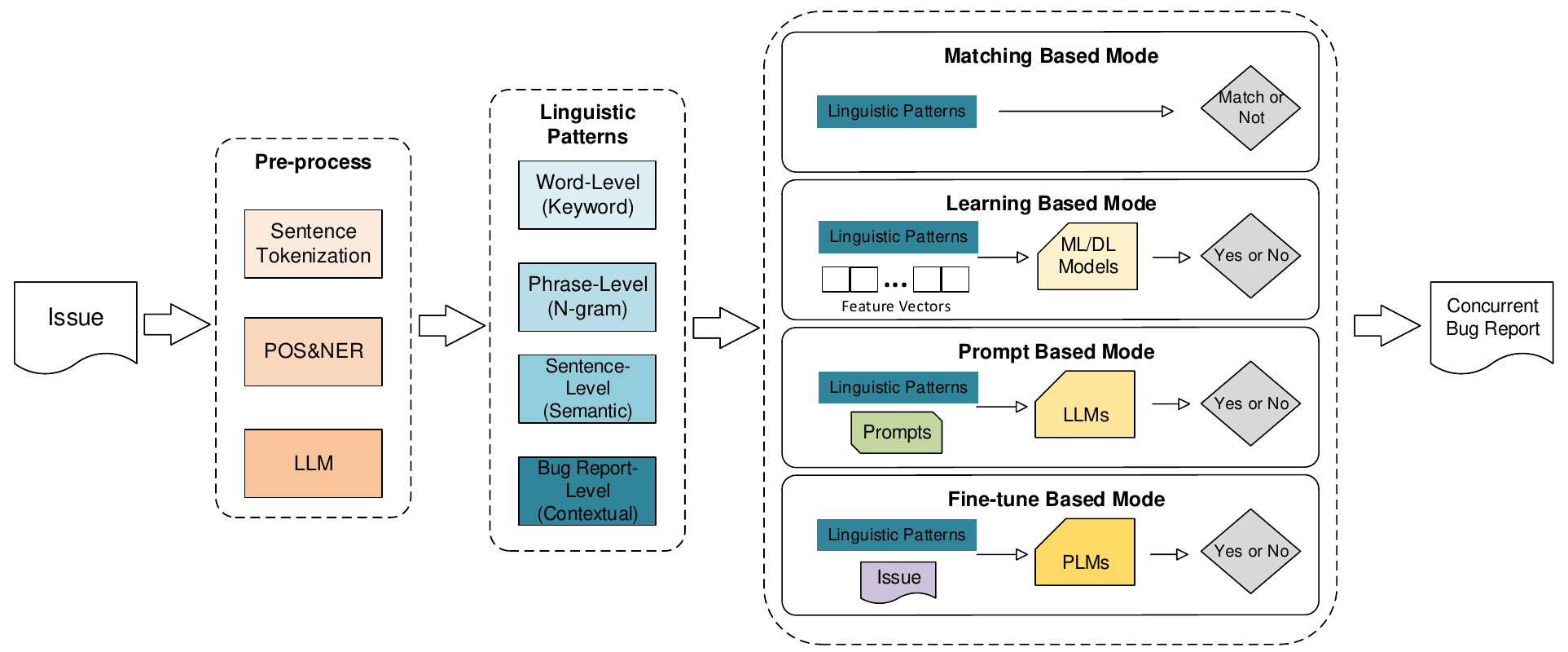}
\caption{Overview of \Name{}} 
\label{frame}
\end{figure*}

Figure~\ref{frame} presents our concurrency bug report identification framework, which consists of pre-processing, linguistic pattern extraction, and classification. Given an issue report, we first tokenize it into sentences and apply Part-of-Speech (POS) tagging, Named Entity Recognition (NER), and LLM-based analysis to extract linguistic patterns at four levels: word-level (keywords), phrase-level (N-grams), sentence-level (semantic), and bug report-level (contextual information). These patterns are then used in four classification approaches: matching-based, which directly checks for pattern matches; learning-based, which converts patterns into feature vectors for ML/DL models; prompt-based, which leverages LLMs for classification; and fine-tune-based, which refines pre-trained language models (PLMs) using labeled bug reports. By integrating linguistic patterns with multiple classification strategies, our framework enhances the accuracy and scalability of concurrency bug report identification.

\subsubsection{Matching-based Approach}

The matching-based approach determines whether any sentences in a bug report match the predefined linguistic patterns (LPs).  
We use LPs across four levels—word, phrase, sentence, and bug report—as search templates.  
If a bug report contains at least one matching pattern, it is classified as a concurrency-related bug report.  
This approach is particularly valuable for its high recall and cost-effectiveness, as it enables fast large-scale screening without the need for model training.

As a fast and efficient mechanism, the matching-based approach can quickly sift through large volumes of bug reports, accurately identifying most concurrency-related issues by matching sentences with predefined LPs. This computationally cost-effective solution ensures broad coverage of potential concurrency bug reports in the dataset, making it an ideal choice for scenarios where achieving high recall is prioritized.

\subsubsection{Learning-based Approach}

The learning-based approach transforms LPs into explicit, interpretable features for traditional machine learning classifiers. 
Each bug report is encoded as a binary feature vector \( v \in \{0,1\}^N \), where \( N \) represents the total number of linguistic patterns (LPs). 
A feature value of 1 indicates the presence of the corresponding LP within the report, enabling models to learn from structured, domain-specific cues rather than relying solely on raw lexical signals. 

We implemented a set of representative learning-based classifiers, including logistic regression, support vector machines, random forests, and gradient-boosted trees, using modern machine learning frameworks with default hyperparameters. 
The use of default configurations ensures unbiased and reproducible baseline results, consistent with established practices in software analytics research~\cite{xia2014automated, panichella2015can}. 
This setup isolates the effect of LP-based features from model-specific tuning, allowing us to directly assess their representational contribution. 
While these models are not optimized for maximum performance, they achieve competitive accuracy and confirm that LPs provide informative and interpretable signals even within standard supervised learning pipelines.

\subsubsection{Prompt-based Approach}

With the emergence of large language models (LLMs), their powerful language understanding and reasoning capabilities have brought new opportunities to automated bug report classification. 
However, these models are not trained on domain-specific corpora such as concurrency bug reports, which limits their accuracy and often leads to hallucination—incorrectly inferring concurrency-related causes when none exist (as demonstrated in Section~\ref{sec:motivation}). 
To address this limitation, we design a prompt-based approach that explicitly injects domain knowledge into LLMs through linguistic patterns (LPs), thereby constraining their reasoning and improving classification consistency.

In this design, LPs act as structured, domain-grounded exemplars that guide the model’s interpretation of concurrency-related concepts. 
We organize LPs hierarchically across four levels (word, phrase, sentence, and bug report) and incorporate them as few-shot examples within the prompt, providing the LLM with progressively richer contextual cues. 
This design mitigates over-generalization and reduces hallucinated inferences by anchoring the model to verified concurrency-related expressions.

An example of the prompt template is shown below:

\begin{tcolorbox}[width=0.9\textwidth, boxrule=0.5pt, colback=white]
\textit{Instruction: Follow the given linguistic patterns as reference examples. 
Analyze the provided bug report and determine whether it describes a concurrency bug. 
Base your reasoning on the relationship between the report content and the patterns.}\\[2mm]
\texttt{[pattern:word]} lock \\
\texttt{[pattern:phrase]} (lock, require) \\
\texttt{[pattern:sentence]} Thread requires a lock \\
\texttt{[pattern:bug report]} Root cause: lock issue \\
\texttt{[bug report]} Thread fails to acquire a lock ... \\
\texttt{[Concurrent bug or not]}: Yes
\end{tcolorbox}

Each prompt begins with LP exemplars labeled by type (\texttt{[pattern:word]}–\texttt{[pattern:bug report]}) and concludes with the target bug report to be classified. 
This hierarchical structure helps the LLM learn the semantic relationship between concurrency-related terms and their contextual roles. 
By including multiple levels of patterns, the model can distinguish superficial mentions of concurrency terms from true concurrency issues.

All prompts are constructed following a few-shot paradigm, where only a small number of LP exemplars are provided for each category.  
This setup leverages the pre-trained reasoning capabilities of LLMs while constraining them with explicit, structured domain knowledge.  
We intentionally adopt few-shot prompting rather than more complex strategies (e.g., Chain-of-Thought or Reason-Then-Act),  
as our goal is to isolate and evaluate the contribution of linguistic patterns themselves,  
rather than the effects of advanced prompt engineering techniques.  
To ensure consistency and reproducibility, we fix the prompt format across all runs and randomize exemplar order within each category.

\subsubsection{Fine-Tuning-based Approach}

While prompt-based classification can effectively leverage linguistic patterns to guide reasoning, it remains sensitive to prompt formulation and often suffers from instability or hallucination in domain-specific contexts. 
To overcome these limitations, we fine-tune pre-trained language models (PLMs) using LPs as structured supervision signals, enabling the models to internalize concurrency-related linguistic knowledge rather than relying solely on contextual prompting.

In this approach, each input to the PLM concatenates the extracted linguistic patterns with the corresponding bug report text. 
This design explicitly embeds domain-specific cues into the training data, allowing the model to focus on features that are semantically and syntactically indicative of concurrency issues. 
An example fine-tuning input format is shown below:

An example fine-tuning input format is shown below:

\begin{tcolorbox}[width=0.9\textwidth, boxrule=0.5pt, colback=white]
\texttt{[CLS]} \\
\texttt{[PATTERN:WORD]} lock \\
\texttt{[PATTERN:PHRASE]} (lock, require) \\
\texttt{[PATTERN:SENTENCE]} Thread requires a lock \\
\texttt{[PATTERN:BUG REPORT]} Root cause: lock issue \\
\texttt{[BUG REPORT]} Thread fails to acquire a lock ... \\
\texttt{[SEP]} \\
\texttt{[LABEL]} concurrency\_bug = 1
\end{tcolorbox}

By providing both the LP exemplars and the raw report text, the model learns to attend to verified concurrency indicators and their contextual manifestations. 
This design serves two complementary purposes: 
(1) it grounds the model’s predictions in explicit, human-validated linguistic evidence, reducing hallucination and false positives, and 
(2) it enhances the stability of predictions by minimizing the sensitivity to prompt phrasing or token ordering.

We fine-tune multiple PLMs, including CodeBERT, GraphCodeBERT, CodeT5, and Codex, to evaluate the generality of our approach. 
Each model is initialized with publicly available pre-trained weights and configured for binary classification. 
The fine-tuning process uses a binary cross-entropy loss function, optimizing over parameters such as learning rate, batch size, and epoch count via grid search to achieve stable convergence. 
During training, LPs and bug report texts are concatenated with special separator tokens, and the model learns to associate these structured representations with the corresponding concurrency labels. 
Early stopping and stratified validation are applied to prevent overfitting.

Fine-tuning PLMs with LP supervision allows the model to internalize domain knowledge beyond surface-level co-occurrence. 
Compared to prompt-based reasoning, this approach achieves higher consistency and generalization by directly embedding concurrency-specific representations into the model parameters. 
As a result, the fine-tuned model demonstrates greater robustness in unseen project data and reduced variance across runs. 

The resulting fine-tuned model, denoted as \Name{}, serves as our final automated system for identifying concurrency bug reports. 
By combining the deep contextual understanding of PLMs with the structured domain guidance of linguistic patterns, \Name{} provides an accurate, scalable, and hallucination-resistant solution for real-world concurrency bug classification.

\subsection{Model Selection}
\label{base}

To evaluate our linguistic patterns, we select six ML/DL models, four LLMs, and five PLMs for concurrency bug report classification, as shown in Table~\ref{tab:models}. ML/DL models use traditional NLP features like TF-IDF and Word2Vec, LLMs rely on prompt-based classification without fine-tuning, and PLMs are fine-tuned on bug report datasets to learn domain-specific patterns. To enhance classification performance, we explore different combinations of our four levels of linguistic patterns across these methods, analyzing how linguistic patterns improve identification accuracy in each category.

\begin{table}[h!] \scriptsize
\centering
\caption{Selected Models.}
\label{tab:models}
\begin{tabularx}{\textwidth}{c>{\centering\arraybackslash}X}
\toprule
\multicolumn{2}{c}{\textbf{Machine/Deep Learning Models}} \\
\midrule
\textit{Models} & \textit{NLP features} \\
\midrule
Naive Bayes (NB) & \\
Logistic Regression (LR) & \\
Support Vector Machine (SVM) & Word Level TF-IDF (TI) \\
Random Forest (RF) & Word2Vec \\
Convolutional Neural Network (CNN) & \\
Long Short-Term Memory (LSTM) & \\
\midrule
\multicolumn{2}{c}{\textbf{Large Language Model}} \\
\midrule
\textit{Model} & \textit{Feature} \\
\midrule
GPT-3.5-turbo & \\
GPT-4o &  \\
GPT-5 & Prompt LLMs to identify concurrency bug reports \\
Claude Sonnet 4.5 & \\
Claude Opus 4.1 & \\
\midrule
\multicolumn{2}{c}{\textbf{Pre-trained Language Model}} \\
\midrule
\textit{Methods} & \textit{Keywords} \\
\midrule
Bert & \\
RoBerta & \\
Albert & Fine tune PLMs using bug report \\
CodeBert & \\
GraphCodeBert & \\
\bottomrule
\end{tabularx}
\end{table}

\subsubsection{Machine/Deep Learning Models} 

Traditional machine and deep learning models are used as baselines to evaluate whether linguistic patterns (LPs) provide information beyond surface-level lexical features. 
We employ six representative models spanning both shallow and neural architectures, including Naïve Bayes (NB), Logistic Regression (LR), Support Vector Machine (SVM), Random Forest (RF), Convolutional Neural Network (CNN), and Long Short-Term Memory (LSTM).  

For feature representation, we use two standard text embedding methods widely adopted in software analytics research: 
TF-IDF for shallow learners (NB, LR, SVM, RF), and Word2Vec for neural models (CNN, LSTM). 
TF-IDF encodes term frequency statistics and remains an effective baseline for text classification tasks due to its simplicity and interpretability. 
Word2Vec generates dense vector embeddings that capture local semantic associations among words, providing richer contextual cues for neural networks.

All models are implemented using modern machine learning frameworks (e.g., scikit-learn, PyTorch) with default hyperparameters to ensure unbiased and reproducible baselines. 
These baselines allow us to isolate the effect of LP-based supervision—demonstrating whether domain-specific linguistic knowledge improves classification accuracy compared to purely statistical features.

\subsubsection{Large Language Models}

We evaluate five state-of-the-art large language models (LLMs): \textbf{GPT-3.5-turbo}, \textbf{GPT-4o}, \textbf{GPT-5}, \textbf{Claude Sonnet 4.5}, and \textbf{Claude Opus 4.1}, representing the latest generations of OpenAI and Anthropic models. 
GPT-5, the most advanced among them, demonstrates significant improvements in reasoning depth, factual grounding, and long-context comprehension, making it particularly effective for domain reasoning tasks such as bug report analysis. 
GPT-4o balances reasoning accuracy and efficiency, while GPT-3.5-turbo serves as a strong lightweight baseline for cost-effective inference. 
Claude Sonnet 4.5 and Claude Opus 4.1, known for their stability and interpretability, provide an additional cross-model perspective beyond the OpenAI ecosystem.

All models are evaluated in a prompt-based setting, leveraging their in-context learning (ICL) capabilities to identify concurrency bug reports without explicit fine-tuning. 
In this setup, linguistic patterns (LPs) are incorporated into the prompt as structured domain cues, allowing LLMs to ground their reasoning in concurrency-specific terminology rather than relying solely on surface-level lexical matches. 
This design mitigates hallucinations and encourages consistent reasoning by constraining the model’s output space to semantically validated LP templates (see Section~\ref{sec:IV}). 

By comparing multiple LLM families and versions, we aim to assess the generalization and domain adaptability of frontier reasoning models when guided by structured linguistic patterns, offering insight into the trade-offs between contextual inference (LLMs) and parameter adaptation (PLMs).

\subsubsection{Pre-trained Language Models}

To further investigate the effect of parameter adaptation and domain grounding, we fine-tune several widely adopted pre-trained language models (PLMs), including \textbf{BERT}, \textbf{RoBERTa}, \textbf{ALBERT}, \textbf{CodeBERT}, and \textbf{GraphCodeBERT}. 
BERT and RoBERTa represent general-purpose transformer architectures capable of capturing contextual dependencies in natural language text, whereas ALBERT provides a lighter, parameter-efficient variant with comparable representational strength.  

For software engineering and code-related contexts, we include CodeBERT and GraphCodeBERT, which are pre-trained on large-scale code repositories and paired NL–PL datasets. 
CodeBERT effectively models both syntactic and semantic relationships between textual descriptions and code elements, while GraphCodeBERT enhances this by encoding program structure through data-flow graphs, yielding improved representations for code semantics.  

Unlike LLMs that perform \textit{contextual inference} through in-context learning, PLMs rely on \textit{parameter adaptation} via fine-tuning on labeled bug report datasets. 
In our setup, each fine-tuning input integrates linguistic patterns (LPs) and the corresponding bug report text, allowing the model to internalize domain-specific cues about concurrency mechanisms and synchronization semantics. 
This design enables PLMs to explicitly learn the mapping between concurrency-related expressions and their classification labels, providing stronger domain alignment than prompt-based inference.  

These fine-tuned PLMs serve as the backbone for our supervised learning experiments, offering a direct comparison against LLM-based prompt reasoning. 
They not only establish a robust baseline for domain-adapted classification but also demonstrate how linguistic patterns enhance model interpretability and stability in concurrency bug report identification.

\section{Evaluation}
\label{sec:evaluation}

We conduct a comprehensive evaluation to assess the effectiveness and generalizability of our linguistic patterns (LPs) and their integration into various classification models. 
Our study addresses the following research questions (RQs):

\vspace*{3pt}
\noindent
\textbf{RQ1: Is the set of linguistic patterns saturated from our manual derivation?}  
This RQ evaluates whether our LPs sufficiently capture the linguistic characteristics of concurrency bug reports and whether adding more data introduces new patterns.

\vspace*{3pt}
\noindent
\textbf{RQ2: How effective are our linguistic patterns in classifying concurrency bug reports?}  
This RQ examines the overall performance of LP-based classification and assesses whether linguistic cues alone can accurately distinguish concurrency bugs.

\vspace*{3pt}
\noindent
\textbf{RQ3: How do different classification approaches leverage linguistic patterns?}  
This RQ compares the performance of four modeling paradigms—matching-based, learning-based, prompt-based, and fine-tuning-based—to analyze how each benefits from LP integration.

\vspace*{3pt}
\noindent
\textbf{RQ4: How do different levels of linguistic patterns contribute to classification performance?}  
This RQ investigates the contribution of LPs at different levels (word, phrase, sentence, and bug report) through ablation analysis to understand their impact on accuracy and interpretability.

\subsection{Dataset Selection}
\label{select}

We evaluate our approach using three datasets constructed from both GitHub and Jira issue tracking systems, as summarized in Table~\ref{evadata}.

\begin{table}[h]
\caption{Evaluation Datasets}\label{evadata}
\centering
\begin{tabular}{|l|c|c|c|}
\hline
\textbf{Dataset} & \textbf{\#Issues} & \textbf{\#ConBugs} & \textbf{\#ConSents} \\ \hline
Dataset$_{Git}$  & 3,640  & 182 & 817 \\ \hline
Dataset$_{Jira}$ & 3,640  & 182 & 751 \\ \hline
Dataset$_{Post}$ & 3,640  & 182 & 682 \\ \hline
\end{tabular}
\end{table}

\begin{itemize}
    \item \textbf{Dataset$_{Git}$.}  
    This dataset is a subset of our linguistic pattern derivation corpus described in Section~\ref{mt}. 
    It originates from 12 GitHub projects comprising 30,982 bug reports, among which 912 are labeled as concurrency-related. 
    We used 80\% (730) of these reports for linguistic pattern derivation, leaving 182 reports as unseen concurrency bug reports for evaluation. 
    To simulate real-world conditions—where concurrency bugs typically account for about 5\% of total bugs~\cite{asadollah2017concurrency}—we randomly sampled 3,458 non-concurrency bug reports from the same pool. 
    Combining the 182 concurrency and 3,458 non-concurrency reports yields 3,640 total issues, ensuring a realistic and balanced benchmark.

    \item \textbf{Dataset$_{Jira}$.}  
    This dataset is constructed from five popular open-source projects on the Jira platform: \textit{Camel}, \textit{Cassandra}, \textit{Hadoop}, \textit{HBase}, and \textit{Hive}. 
    It contains 3,640 issues in total, including 182 concurrency-related reports and 3,458 non-concurrency reports. 
    We ensure that Dataset$_{Jira}$ does not overlap with Dataset$_{Git}$ to maintain a strict separation between training and evaluation data. 
    This allows us to evaluate the generalizability of our approach on unseen projects and issue tracking systems.

    \item \textbf{Dataset$_{Post}$.}  
    To assess robustness and forward-transfer capability, we build a post-cutoff dataset composed of bug reports submitted \emph{after} the training data cutoff of most large language models (October 2023). 
    Dataset$_{Post}$ includes reports from the same GitHub and Jira projects as Dataset$_{Git}$ and Dataset$_{Jira}$, collected exclusively from issues created after October 2023. 
    This dataset serves as a testbed for evaluating whether models—especially LLM-based approaches—can generalize to newly reported, previously unseen bugs.
\end{itemize}

Overall, our evaluation covers 10,920 bug reports across 12 projects, including 546 labeled concurrency bug reports. 
This multi-source setup ensures a comprehensive and realistic assessment of our linguistic-pattern-based classification framework under diverse data conditions.

\subsection{Results Analysis}

\subsubsection{RQ1: Is the set of linguistic patterns saturated from our manual derivation?}

We determine the "saturation" of our linguistic pattern (LP) extraction based on two key criteria: 
\textit{stability over iterations} and \textit{generalizability across projects}. 
For stability over iterations, we observe the evolution of linguistic patterns across successive iterations, 
identifying saturation when the emergence of new LPs halts or slows significantly—indicating that additional iterations are unlikely to yield substantial novel insights. 
The second criterion, generalizability across projects, evaluates whether the extracted LPs can effectively cover concurrency-related bug reports from unseen projects. 
Here, we focus primarily on \textit{recall}, which measures the coverage of existing LPs on new datasets. 
High recall on unseen projects suggests that the derived LPs generalize well beyond the original training corpus. 
This generalizability indicates that our LP set has reached a stable and representative state, ensuring its robustness and practical applicability for concurrency bug identification across diverse software systems.
We use 80\% of concurrency bug reports (730) from our dataset, which contain 2,553 concurrency related sentences, to derive linguistic patterns (LPs).

\noindent
\textbf{Issue-wise Iterative Saturation.}
To examine how LPs evolve as more bug reports are processed, 
we randomly divide the 2,553 concurrency-related sentences into ten subsets 
and iteratively extract LPs by adding one subset at a time. 

\begin{figure}[htbp]
	\centering
	\begin{minipage}{0.49\linewidth}
		\includegraphics[width=\linewidth]{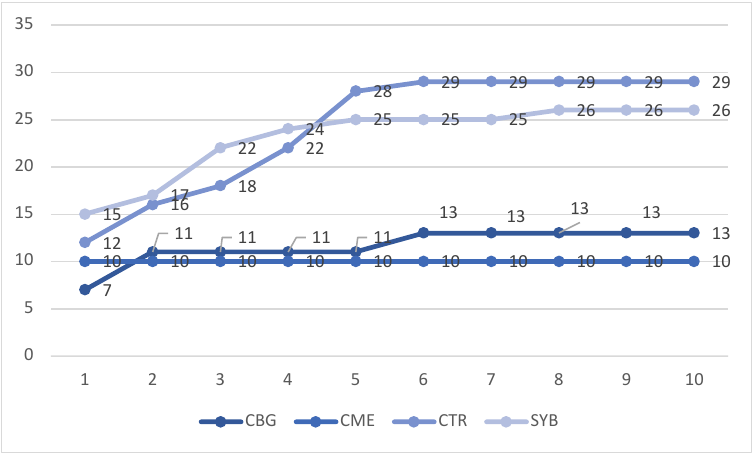}
		\subcaption{Iterations of NOUN.}
	\end{minipage}
	\begin{minipage}{0.49\linewidth}
		\includegraphics[width=\linewidth]{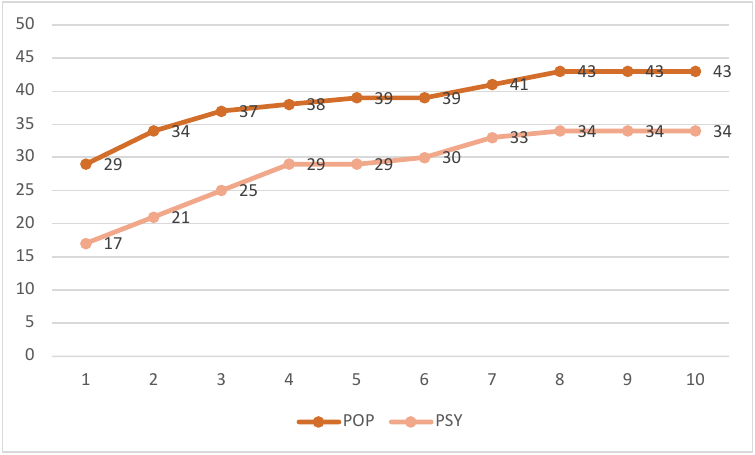}
		\subcaption{Iterations of VERB.}
	\end{minipage}
    \\
    \begin{minipage}{0.49\linewidth}
		\includegraphics[width=\linewidth]{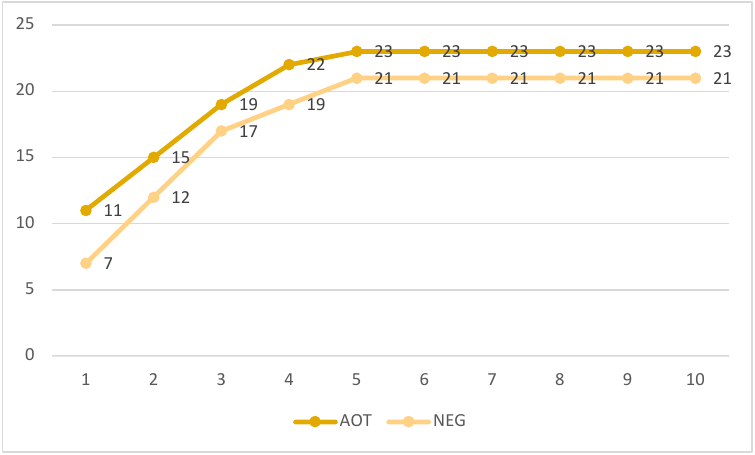}
		\subcaption{Iterations of ADV.}
	\end{minipage}
	\begin{minipage}{0.49\linewidth}
		\includegraphics[width=\linewidth]{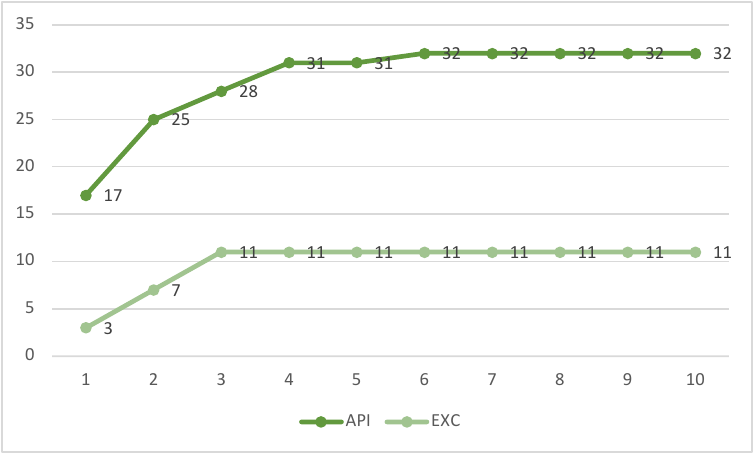}
		\subcaption{Iterations of API.}
	\end{minipage}
	\caption{Iterations of Different Types of Word.}
	\label{iterword}
\end{figure}

Figure~\ref{iterword} illustrates the growth trends of four word categories: 
nouns (CBG, CME, CTR, SYB), verbs (POP, PSY), adverbs (AOT, NEG), and API-related terms (API, EXC). 
Most categories exhibit rapid expansion within the first three to four iterations, 
followed by a clear plateau. 
For instance, noun categories reach stability by iteration~6 (CBG~=~13, CME~=~10, CTR~=~29, SYB~=~26), 
verbs by iteration~8 (POP~=~43, PSY~=~34), and API-related terms even earlier (API~=~32, EXC~=~11). 
This convergence suggests that the majority of concurrency-related lexical cues 
are captured within the early extraction cycles.

\begin{figure}[htbp]
	\centering
	\begin{minipage}{0.49\linewidth}
		\includegraphics[width=\linewidth]{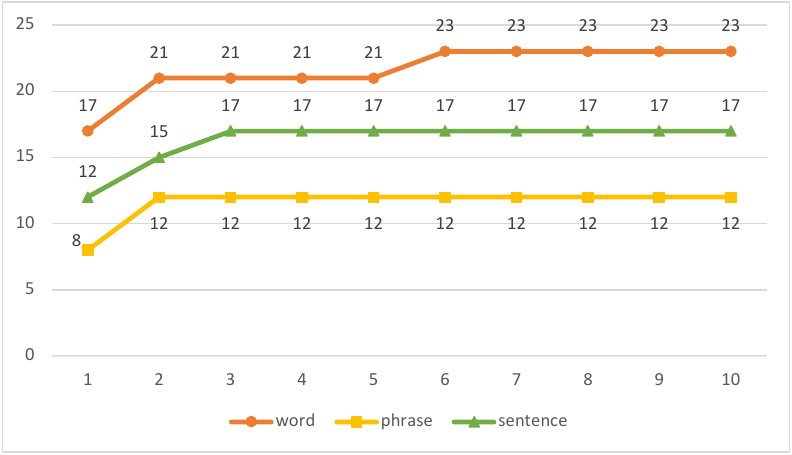}
		\subcaption{The Number of Patterns with Iterations.}
	\end{minipage}
	\begin{minipage}{0.49\linewidth}
		\includegraphics[width=\linewidth]{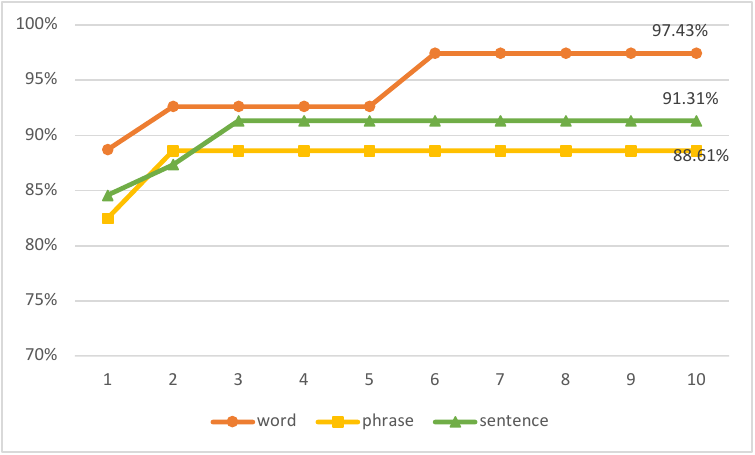}
		\subcaption{The recall of patterns with iteration.}
	\end{minipage}
	\caption{Linguistic Patterns Saturation with Iterations.}
	\label{lp}
\end{figure}

For linguistic patterns (Figure~\ref{lp}), 
word-level LPs grow from~17 to~23 by iteration~6 and then plateau; 
phrase- and sentence-level LPs converge at~12 and~17, respectively.  
Bug report-level LPs (e.g., \textit{lock issue}, \textit{thread symptom}) remain constant throughout, 
as they represent stable semantic templates describing the overall context of a bug report 
rather than individual concurrency-related sentences. 
We do not explicitly show the growth curve for bug report-level LPs 
because they are derived at the report level—essentially an extension of the sentence-level LPs. 
Thus, once sentence-level patterns approach saturation, 
bug report-level patterns have already stabilized or become saturated earlier.  
Recall on a held-out set of 817 concurrency-related sentences shows a similar trend:  
word-level recall increases from~88.7\% to~92.6\%, phrase-level from~82.5\% to~88.6\%, 
and sentence-level reaches~91.3\% by iteration~6, after which all curves flatten.  
This pattern indicates that linguistic diversity in concurrency bug descriptions 
is largely captured within a small number of iterations.

\noindent
\textbf{Project-wise Iterative Saturation.}
To evaluate cross-project generality, 
we extend LP extraction sequentially across seven representative projects ordered by report count.

\begin{figure}[htbp]
	\centering
	\begin{minipage}{0.49\linewidth}
		\includegraphics[width=\linewidth]{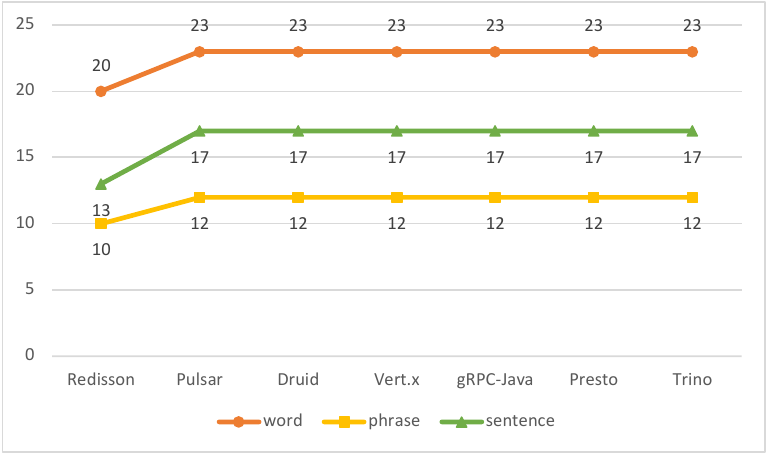}
		\subcaption{The Number of Patterns with Iterations.}
	\end{minipage}
	\begin{minipage}{0.49\linewidth}
		\includegraphics[width=\linewidth]{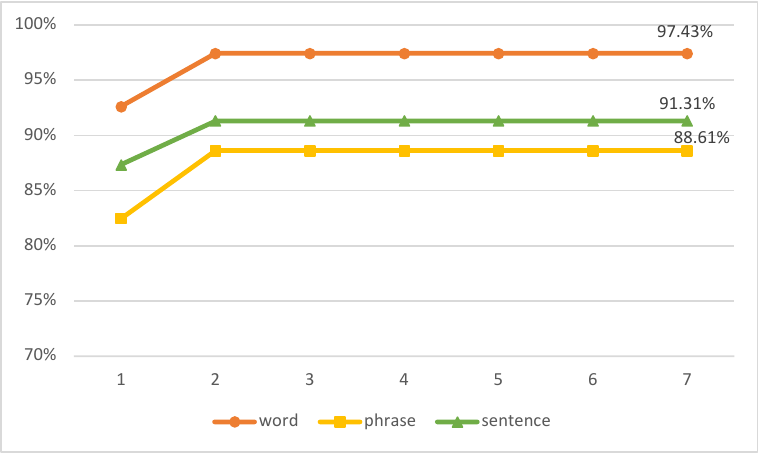}
		\subcaption{The recall of patterns with iteration.}
	\end{minipage}
	\caption{Linguistic Patterns Saturation by Projects.}
	\label{plp}
\end{figure}

Figure~\ref{plp}(a) shows that most new LPs appear within the first two projects (\textit{Redisson} and \textit{Pulsar}), 
after which the total number of LPs remains constant 
(23 word-level, 12 phrase-level, and 17 sentence-level).  
Figure~\ref{plp}(b) shows that recall similarly stabilizes after two projects 
(97.4\%, 88.6\%, and 91.3\% for word-, phrase-, and sentence-level LPs, respectively).  
This rapid convergence suggests that concurrency-related language exhibits strong inter-project regularity, 
and a small set of representative projects suffices to cover most linguistic variations.

\noindent
\textbf{Cross-Dataset Generalization.}
Finally, we assess LP generalizability on two unseen datasets—Dataset$_{Jira}$ and Dataset$_{Post}$—
containing 751 and 682 concurrency-related sentences, respectively.

\begin{figure}[htbp]
	\centering
	\begin{minipage}{0.49\linewidth}
		\includegraphics[width=\linewidth]{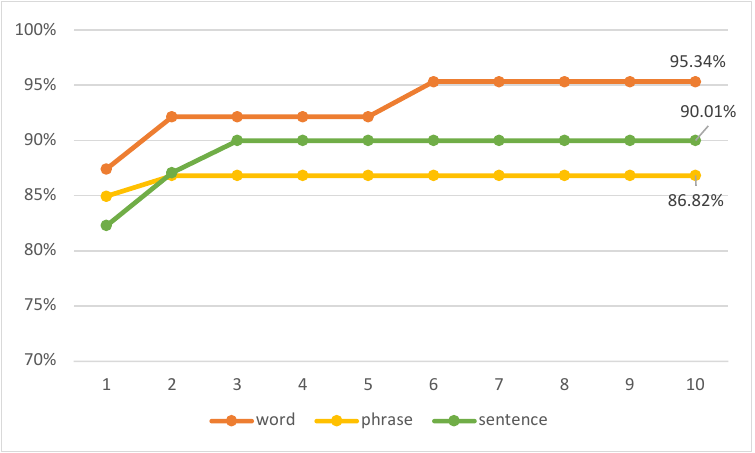}
		\subcaption{Recall on Dataset$_{Jira}$.}
	\end{minipage}
	\begin{minipage}{0.49\linewidth}
		\includegraphics[width=\linewidth]{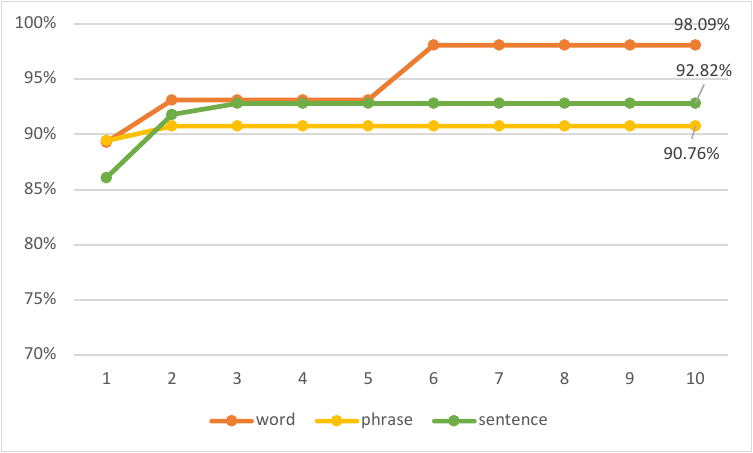}
		\subcaption{Recall on Dataset$_{Post}$.}
	\end{minipage}
	\caption{Cross-Dataset Generalization of Linguistic Patterns.}
	\label{lpgj}
\end{figure}

As shown in Figure~\ref{lpgj}, the LPs maintain consistently high recall across datasets.  
On Dataset$_{Jira}$, recall reaches 95.34\%, 86.82\%, and 90.01\% for word-, phrase-, and sentence-level patterns, respectively.  
On Dataset$_{Post}$, recall slightly improves to 98.09\%, 92.82\%, and 90.76\%,  
confirming that the LPs not only generalize across projects but also remain temporally stable.  
Together with the project-wise analysis, this provides strong evidence that our LPs capture 
fundamental, reusable linguistic structures of concurrency bug reports.

\vspace*{3pt}
\begin{tcolorbox}[width=\textwidth, boxrule=0.5pt, colback=gray!30]
{\bf Answer to RQ1:}  {
Across issue-wise, project-wise, and cross-dataset analyses,  
our LP extraction consistently converges within six iterations or two projects,  
while maintaining high recall on unseen datasets.  
These findings confirm that the derived LPs are both \textit{saturated} and \textit{generalizable}, 
forming a stable and comprehensive linguistic foundation for downstream concurrency bug classification.
}
\end{tcolorbox}

\subsubsection{RQ2: How effective are our linguistic patterns in classifying concurrency bug reports?}

We apply the matching-based approach on Dataset$_{Git}$ using four types of linguistic patterns (LPs): word-level, phrase-level, sentence-level, and bug report-level (BR). 
Table~\ref{match} summarizes the corresponding precision, recall, and F-measure.

Word-level LPs achieve high recall (0.98) but very low precision (0.12), indicating broad coverage but substantial noise, which results in a low F-measure (0.21).  
Phrase-level LPs slightly improve precision (0.15) while maintaining high recall (0.86), suggesting that local contextual cues help reduce false positives.  
Sentence-level LPs further enhance performance, reaching 0.29 precision, 0.85 recall, and an F-measure of 0.43, showing that semantic context contributes to more accurate matches.  
Finally, bug report-level (BR) LPs yield the best overall balance, with 0.69 precision, 0.70 recall, and an F-measure of 0.69.  
In summary, lower-level LPs (word and phrase) provide broad lexical coverage but are prone to false positives due to their inability to capture semantic intent.  
In contrast, higher-level LPs (sentence and bug report) leverage contextual semantics and discourse-level cues, leading to more accurate and explainable identification of concurrency bug reports.

\begin{table}[]
    \centering
    \caption{Matching-Based Approach on Dataset$_{Git}$.}\label{match}
    \begin{tabular}{ccccc}
    \toprule[1pt]
    Dataset   & Method & Precision & Recall & F-Measure \\ \midrule
    \multirow{4}{*}{Dataset$_{Git}$} & Word         & 0.12    & \textbf{0.98}  & 0.21     \\
                                     & Phrase       & 0.15    & 0.86  & 0.25      \\
                                     & Sentence     & 0.29    & 0.85  & 0.43      \\
                                     & Bug report   & \textbf{0.69}    & 0.70  & \textbf{0.69}    \\ \bottomrule[1pt]
    \end{tabular}
    \end{table}

Our learning-based approach includes two categories of models: machine learning (ML) and deep learning (DL). 
Linguistic patterns (LPs) at four levels—word (KW), phrase (PH), sentence (SE), and bug report (BR)—are transformed into binary feature vectors.  
We further explore various feature combinations (e.g., KW+PH, KW+PH+SE+BR) to evaluate their cumulative effects on classification performance.

\begin{figure*}[htbp]
	\centering
	\begin{minipage}{0.31\linewidth}
		\centering
		\includegraphics[width=1\linewidth]{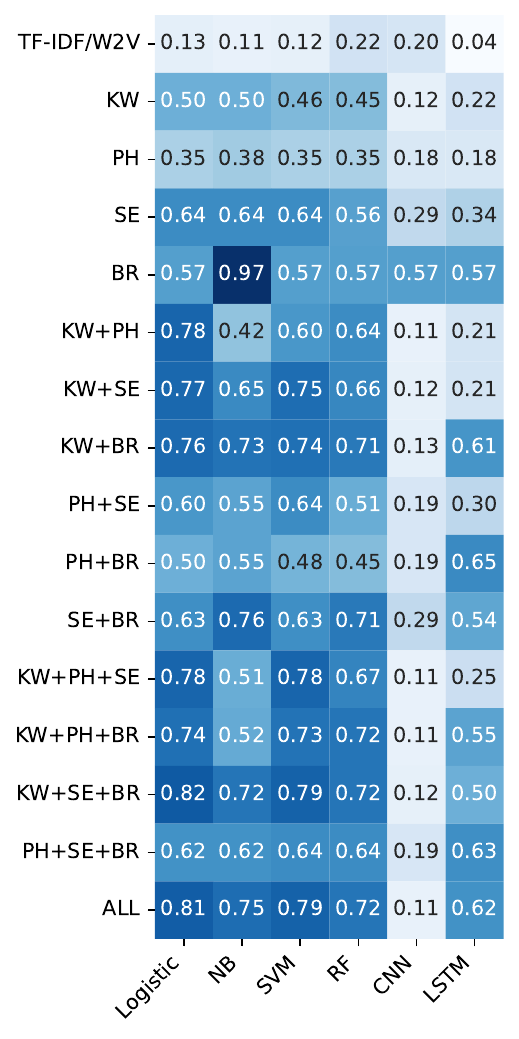}
		\subcaption{Precision.}
		\label{lpa}
	\end{minipage}
	\begin{minipage}{0.31\linewidth}
		\centering
		\includegraphics[width=1\linewidth]{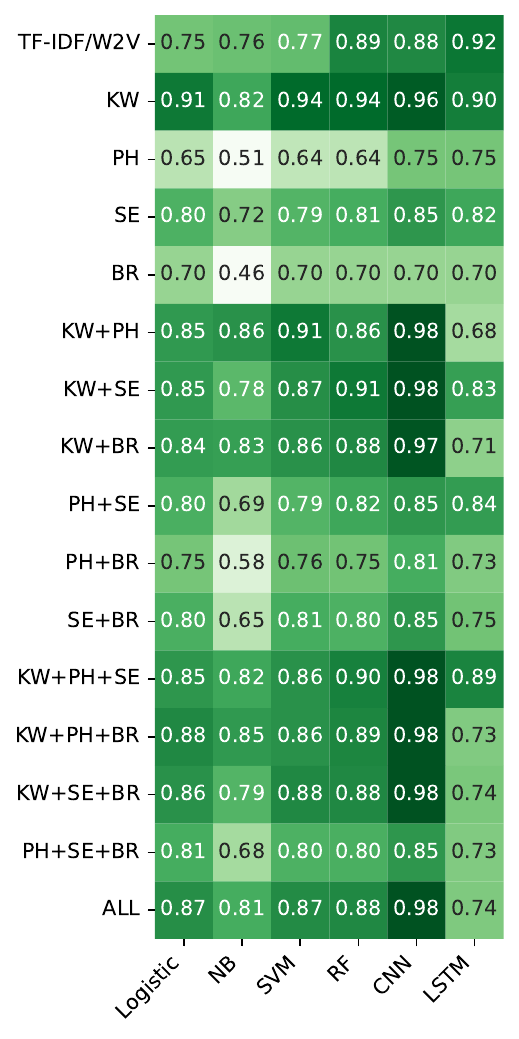}
		\subcaption{Recall.}
		\label{lpb}
	\end{minipage}
    \begin{minipage}{0.31\linewidth}
		\centering
		\includegraphics[width=1\linewidth]{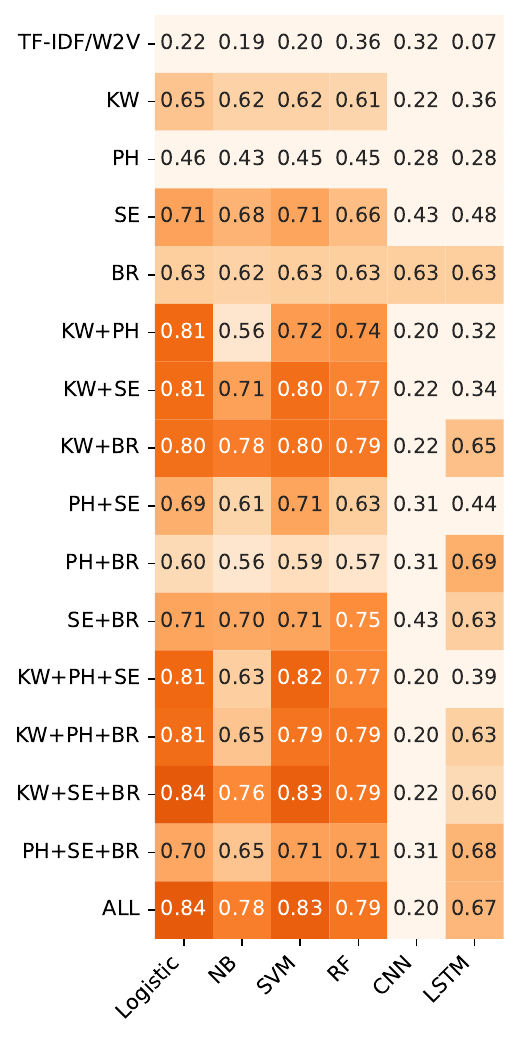}
		\subcaption{F1 score.}
		\label{lpb}
	\end{minipage}
	\caption{ML/DL+linguistic patterns on Dataset$_{Git}$.}
	\label{mllps}
\end{figure*}

Figure~\ref{mllps} presents the precision, recall, and F1-scores of ML and DL models trained with different LP combinations under 10-fold cross-validation on Dataset$_{Git}$.  
Overall, both ML and DL approaches substantially outperform the matching-based baseline, confirming the benefit of supervised learning with LP features.  
Precision improves notably when higher-level LPs are incorporated, with Logistic Regression achieving the best performance—0.81 precision, 0.87 recall, and an F1-score of 0.84 when using all LPs (ALL).  

Among ML models, Logistic Regression and SVM consistently yield the highest and most stable results, while Random Forest performs moderately.  
In contrast, DL models such as CNN and LSTM show weaker and less stable results, with F1-scores ranging between 0.20 and 0.70, reflecting their sensitivity to the limited dataset size and sparse LP feature space.

Combining multiple LPs (e.g., KW+PH or KW+PH+SE+BR) generally boosts recall—for instance, recall rises from 0.70 (BR only) to 0.87 (ALL) for Logistic Regression—indicating broader coverage of concurrency-related expressions.  
However, precision gains are marginal compared to using BR patterns alone, suggesting that excessive feature fusion introduces redundancy rather than new discriminative information.  
This effect is more pronounced in DL models, where adding lower-level patterns offers limited improvement over BR-level features.

In summary, ML/DL-based approaches leveraging linguistic patterns achieve strong and balanced classification performance.  
Lower-level LPs (KW, PH) emphasize recall, whereas higher-level LPs (SE, BR) yield the best precision and overall F1-scores.  
These results demonstrate that semantic and contextual linguistic representations are most effective for distinguishing concurrency-related bug reports in supervised learning frameworks.

\begin{figure}[htbp]
	\centering
	\begin{minipage}{0.18\linewidth}  
		\centering
		\includegraphics[width=1\linewidth]{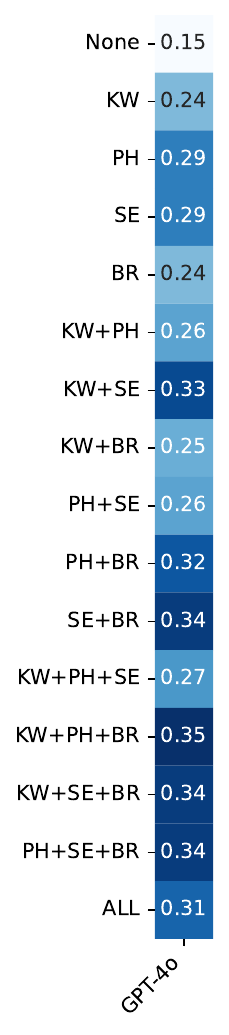}
		\subcaption{Precision.}
		\label{lpa}
	\end{minipage}
	\begin{minipage}{0.18\linewidth}  
		\centering
		\includegraphics[width=1\linewidth]{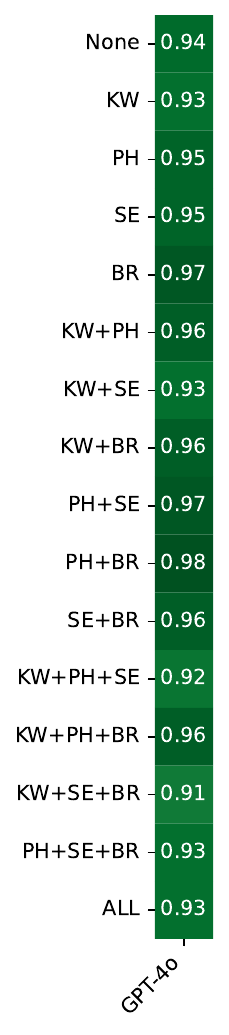}
		\subcaption{Recall.}
		\label{lpb}
	\end{minipage}
    \begin{minipage}{0.18\linewidth}  
		\centering
		\includegraphics[width=1\linewidth]{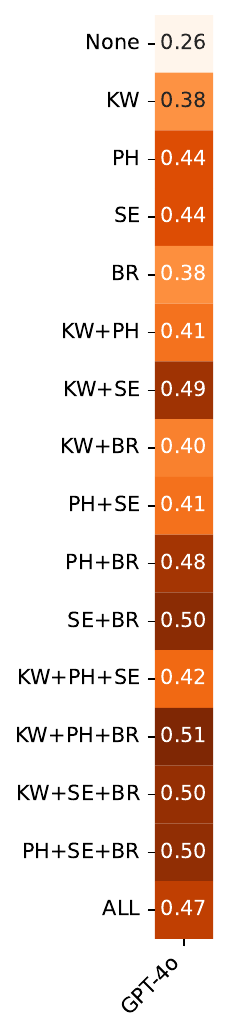}
		\subcaption{F1 score.}
		\label{lpc}
	\end{minipage}
	\caption{GPT+linguistic patterns on Dataset$_{Git}$.}
	\label{gptlps}
\end{figure}

Due to the high computational and financial cost of large language models, we primarily conducted full-scale experiments using \texttt{GPT-4o}, which demonstrated the most consistent performance among tested LLMs. 
Other models were evaluated on partial subsets for comparative analysis, and their results are discussed in RQ3.  

We incorporate linguistic patterns as few-shot examples in the prompt to guide GPT-4o toward recognizing concurrency-related semantics. 
This design provides structured contextual cues to enhance the model’s understanding of domain-specific behaviors. 
As shown in Figure~\ref{gptlps}, the use of LPs yields modest but consistent improvements, especially in precision when sentence-level (SE) and bug report-level (BR) patterns are included. 
However, due to the model’s lack of domain-specific pretraining, the overall F1-scores remain lower than those achieved by ML/DL-based approaches.

In summary, linguistic patterns help \texttt{GPT-4o} better interpret concurrency semantics in few-shot settings, but the absence of fine-tuning on domain data limits its overall performance compared to supervised learning methods.

\begin{figure*}[htbp]
	\centering
	\begin{minipage}{0.31\linewidth}
		\centering
		\includegraphics[width=1\linewidth]{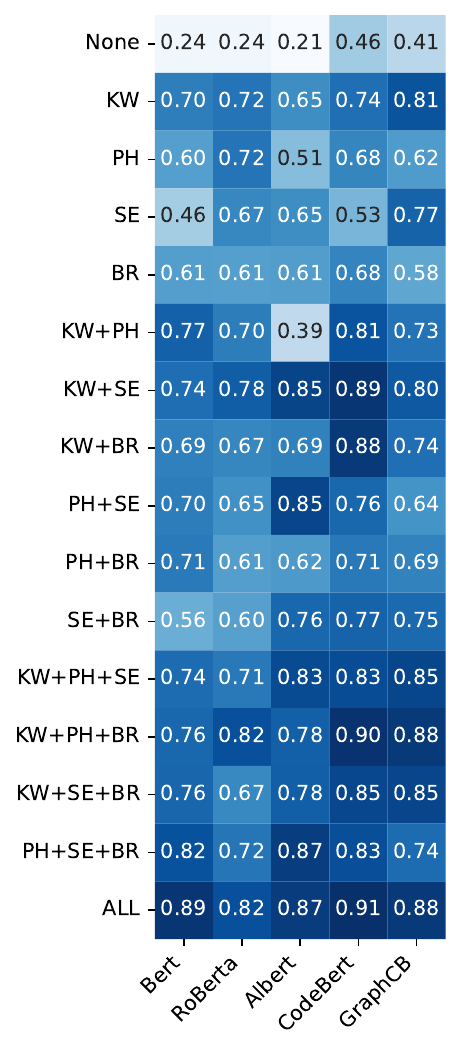}
		\subcaption{Precision.}
		\label{lpa}
	\end{minipage}
	\begin{minipage}{0.31\linewidth}
		\centering
		\includegraphics[width=1\linewidth]{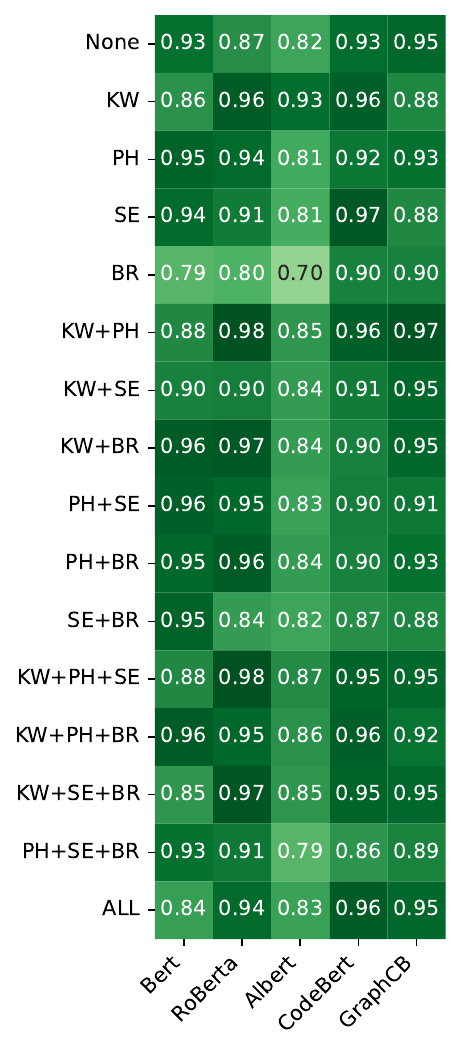}
		\subcaption{Recall.}
		\label{lpb}
	\end{minipage}
    \begin{minipage}{0.31\linewidth}
		\centering
		\includegraphics[width=1\linewidth]{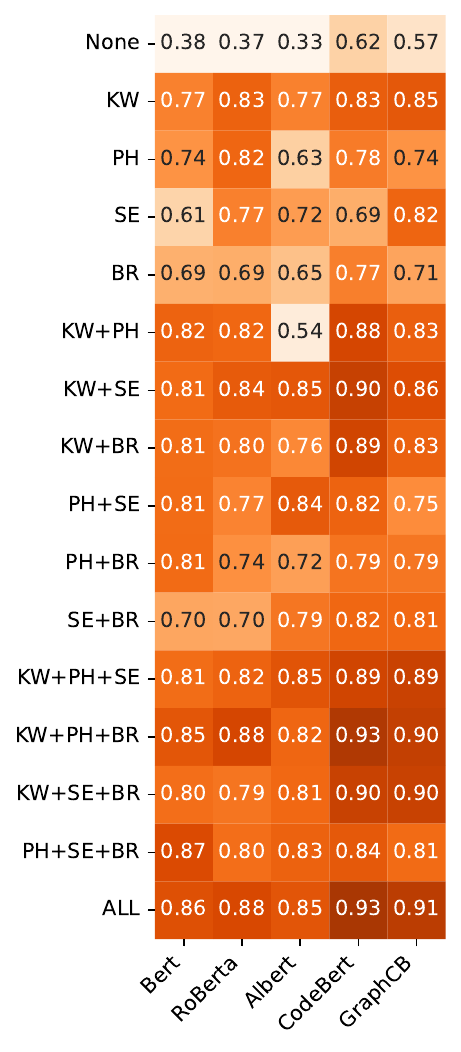}
		\subcaption{F1 score.}
		\label{lpb}
	\end{minipage}
	\caption{PLM+linguistic patterns on Dataset$_{Git}$.}
	\label{plmlps}
\end{figure*}

To evaluate the impact of linguistic patterns (LPs) on fine-tuned pre-trained language models (PLMs), 
we fine-tune five representative models—BERT, RoBERTa, ALBERT, CodeBERT, and GraphCodeBERT—on our dataset.  
Figure~\ref{plmlps} reports the precision, recall, and F1-scores across individual LPs (KW, PH, SE, BR) and their combinations.

Overall, incorporating LPs substantially improves the performance of all PLMs compared to the baseline (None).  
For instance, using all LPs increases the F1-score of ALBERT from 0.33 to 0.83, 
highlighting the strong discriminative power of contextual and semantic cues.  
Combining multiple LPs further boosts recall and overall F1, as broader LP combinations capture a richer spectrum of linguistic signals.  
Among all models, CodeBERT achieves the best performance with an F1-score of 0.93 when using all LPs, while GraphCodeBERT follows closely (F1 = 0.90) even with fewer feature combinations (e.g., KW+SE+BR).  

These results demonstrate that CodeBERT and GraphCodeBERT, pre-trained on code and software-related corpora, consistently outperform general-purpose PLMs, 
suggesting that software-aware representations are inherently better suited for this task.
LP-enriched fine-tuning markedly enhances PLM performance, 
with task-specific models (CodeBERT, GraphCodeBERT) achieving the best balance between precision and recall.  
Among them, CodeBERT combined with all levels of linguistic patterns (\texttt{CodeBERT+ALL}) achieves the highest F1-score (0.93), 
and statistical significance tests (\textit{Wilcoxon signed-rank}, $\alpha=0.05$) confirm that its improvement over other PLMs, 
learning-based, and prompt-based baselines is significant ($p<0.05$).  
We refer to this configuration as \Name{}, our final LP-augmented classification model.  

These findings confirm that linguistic patterns serve as effective domain-level guidance, 
enhancing contextual alignment and reducing ambiguity in model predictions.  
\Name{} thus provides a robust and interpretable framework for concurrency bug report classification, 
combining the strengths of code-aware PLMs with structured linguistic knowledge.

\noindent
\textbf{Cross-Project Evaluation.}
To evaluate the generalizability of our linguistic pattern (LP)-based approaches, 
we conduct cross-project testing on Dataset$_{Jira}$, which includes bug reports from five projects unseen during LP derivation.  
For clarity, we report the best-performing representative from each category: 
the matching-based (Match-BR), learning-based (Logistic-ALL), prompt-based (GPT-4o-ALL), and fine-tuned PLM-based (CodeBERT-ALL) methods.  

As shown in Table~\ref{jira_re}, CodeBERT achieves the highest overall performance (F-measure = 0.95), 
demonstrating excellent cross-project generalization and the strongest alignment between LP-enhanced fine-tuning and domain semantics.  
The logistic regression model also performs competitively (F-measure = 0.85), confirming that LPs serve as reliable and interpretable features even in conventional ML classifiers.  
By contrast, the prompt-based GPT-4o struggles to generalize (F-measure = 0.45), despite using LP-enriched exemplars, reflecting its limited transferability without explicit domain adaptation.  
The matching-based baseline (Match-BR) yields moderate results (F-measure = 0.70), indicating that manually derived LPs retain partial cross-project robustness but lack fine-grained contextual adaptation.
Overall, these results show that LP-augmented fine-tuned PLMs—especially CodeBERT—achieve superior robustness across software domains, 
while LP-guided ML methods remain competitive for interpretable, lightweight classification.  
A detailed per-project breakdown and additional model comparisons are provided in our replication package~\cite{shao2025replication}.

\begin{table}[]
    \centering
    \caption{Overall performances on Dataset$_{Jira}$.}\label{jira_re}
    \begin{tabular}{ccccc}
    \toprule[1pt]
    Dataset   & Method & Precision & Recall & F-Measure \\ \midrule
    \multirow{4}{*}{Dataset$_{Jira}$} & Match (BR)         & 0.71    & 0.69  & 0.70     \\
                                     & Logistic (ALL)      & \textbf{0.93}    & 0.78  & 0.85     \\
                                     & GPT-4o (ALL)   & 0.32    & 0.79  & 0.45      \\
                                     & CodeBert (ALL)   & \textbf{0.93}    & \textbf{0.97}  & \textbf{0.95}    \\ \bottomrule[1pt]
    \end{tabular}
    \end{table}

\noindent
\textbf{Post Cut-off Evaluation.}
We further test temporal robustness using Dataset$_{Post}$, which contains bug reports submitted after the temporal cut-off of all large language models.  

Table~\ref{post_re} shows that both CodeBERT and Logistic achieve the highest F-measure (0.93), 
indicating that LP-enhanced fine-tuning and feature-based learning maintain high stability even under temporal drift.  
However, the learning-based model (e.g., Logistic) is more sensitive to data distribution shifts—its F-measure drops from 0.93 on Dataset$_{Post}$ to 0.85 on Dataset$_{Jira}$— 
whereas CodeBERT remains consistently robust, sustaining an F-measure above 0.91 across both datasets.  

In summary, post cut-off evaluation confirms that LP-enriched models—particularly fine-tuned PLMs—retain strong generalization beyond their training boundary.  
Together with the cross-project results, these findings highlight the robustness, scalability, and practical viability of linguistic pattern–guided learning for real-world concurrency bug classification.  
Comprehensive results and replication materials are available in our public package~\cite{shao2025replication}.

\begin{table}[]
    \centering
    \caption{Overall performances on Dataset$_{Post}$.}\label{post_re}
    \begin{tabular}{ccccc}
    \toprule[1pt]
    Dataset   & Method & Precision & Recall & F-Measure \\ \midrule
    \multirow{4}{*}{Dataset$_{Post}$} & Match (BR)         & 0.67    & 0.75  & 0.71     \\
                                     & Logistic (ALL)      & \textbf{0.94}   & 0.92  & \textbf{0.93}      \\
                                     & GPT-4o (ALL)   & 0.41    & 0.93  & 0.57      \\
                                     & CodeBert (ALL)   & 0.91    & \textbf{0.95}  & \textbf{0.93}    \\ \bottomrule[1pt]
    \end{tabular}
    \end{table}

\vspace*{3pt}
\begin{tcolorbox}[width=\textwidth, boxrule=0.5pt, colback=gray!30]
{\bf Answer to RQ2:}  {
    Our results confirm that linguistic patterns (LPs) substantially enhance the classification of concurrency bug reports across diverse models and datasets. 
Matching-based methods achieve high recall but low precision due to lexical noise, while incorporating LPs as structured features in ML/DL models yields a large improvement in F-measure. 
Fine-tuning pre-trained language models (PLMs) with LP-enriched inputs delivers the highest overall accuracy (F-measure above 0.90), demonstrating that LPs provide domain-specific contextual cues that are effectively captured during fine-tuning. 
Prompt-based LLMs (e.g., GPT-4o) also benefit modestly from LP-guided prompting, though their performance remains constrained without task-specific adaptation.  
Cross-project and post cut-off evaluations further demonstrate the robustness and generality of our LP-based approach across different projects, time periods, and model architectures.
}
\end{tcolorbox}

\subsubsection{RQ3: What is the impact of different approaches in leveraging our linguistic patterns for classifying concurrency bug reports?}

\noindent
\textbf{Matching-Based Approach.}
We first applied a matching-based baseline using different levels of linguistic patterns (LPs) on Dataset$_{Git}$.  
Word-level matching achieved very high recall (0.98) but extremely low precision (0.12), as many concurrency-related keywords (e.g., “thread,” “lock,” “synchronize”) frequently appear in non-concurrency contexts.  
Even when incorporating keyword sets from prior work~\cite{asadollah2017concurrency, abbaspour201710}, precision remained low, confirming that purely lexical cues are insufficient for distinguishing true concurrency semantics.  

Phrase-level matching moderately improved precision (0.15) by capturing short contextual expressions (e.g., “acquire lock,” “thread pool”), yet semantic ambiguity persisted, as such phrases often describe benign operations rather than concurrency faults.  
Sentence-level matching offered clearer gains (F1 = 0.43) by integrating syntactic and semantic dependencies, thus reducing spurious matches from incidental keyword co-occurrences.  
Finally, bug report-level (BR) matching achieved the best overall balance (precision = 0.69, recall = 0.70, F1 = 0.69), benefiting from contextual reasoning at the report level that distinguishes causal concurrency failures from superficial mentions.  

Overall, the results show that increasing linguistic granularity—from individual words to full reports—progressively enhances contextual understanding and mitigates lexical noise.  
However, the limitations of keyword- and phrase-based matching highlight the need for learning-based models that can internalize LP structures and reason about concurrency semantics more robustly.

\noindent
\textbf{Learning-Based Approaches.}
To assess the effectiveness of LPs in supervised learning, we trained six classifiers—Logistic Regression, Naïve Bayes (NB), Support Vector Machine (SVM), Random Forest (RF), Convolutional Neural Network (CNN), and Long Short-Term Memory (LSTM)—using different LP combinations as binary feature vectors.  

Figure~\ref{mllps} shows clear distinctions between model families.  
Among traditional ML models, Logistic Regression achieved the best and most stable performance (F1 = 0.84 with all LPs), while SVM performed comparably, demonstrating that LPs provide discriminative and interpretable features well-suited for linear decision boundaries.  
Random Forest produced moderate but less consistent results, reflecting its limited sensitivity to sparse binary inputs.

In contrast, deep learning models (CNN, LSTM) performed substantially worse (F1 $\leq$ 0.32) due to several factors:  
(1) the sparse, symbolic representation of LPs provides limited utility for models optimized for dense embeddings;  
(2) the dataset size (912 labeled reports) is too small for neural architectures to learn reliable feature hierarchies; and  
(3) LPs already encode high-level semantics, diminishing the benefits of end-to-end representation learning.  
As a result, DL models overfit quickly and fail to generalize across unseen reports.

Overall, ML models exploit LPs more effectively than DL models, achieving high accuracy with minimal data.  
Their explainability and efficiency make them particularly suitable for low-resource domains like concurrency bug classification.

\textbf{Prompt-Based Approaches.}
We next evaluated four large language models (LLMs)—GPT-3.5, GPT-4o, GPT-5, and Claude Sonnet—under several prompting strategies (direct, few-shot, chain-of-thought, and reason-then-act).  
As shown in Table~\ref{dllm}, all models struggled with concurrency bug classification due to the absence of domain-specific pretraining, though GPT-4o achieved the best few-shot performance (F1 = 0.47).  
Incorporating LPs as few-shot exemplars yielded modest precision gains—especially for sentence-level and bug report-level patterns—but results remained below those of fine-tuned PLMs.  
This demonstrates that while LPs provide contextual guidance, prompting alone cannot compensate for the lack of concurrency-aware representations.

Model differences arise from their pretraining and reasoning capacities.  
GPT-3.5 and Claude Sonnet show high recall but poor precision, often over-predicting concurrency involvement by over-relying on lexical cues.  
GPT-5 achieves slightly better reasoning consistency but still lacks exposure to developer-oriented text.  
GPT-4o performs best overall due to its partial exposure to technical and code-related data, allowing better grounding between LP exemplars and concurrency semantics.  
Nevertheless, without fine-tuning, even GPT-4o cannot capture complex causal reasoning like thread interleaving or deadlock detection.

We restricted experiments to the \textit{few-shot} setting to isolate the impact of LPs themselves, avoiding the confounding effects of other prompt-engineering techniques (e.g., CoT, RTA).  
By fixing the prompt structure and varying only LP exemplars, we ensure a controlled comparison of LP-guided prompting versus unguided baselines.

\begin{table}[]
    \centering
    \renewcommand{\arraystretch}{1.1}
    \caption{Different LLMs on Dataset$_{Git}$.}\label{dllm}
    \begin{tabular}{clccc}
    \toprule[1pt]
    Dataset   & Method & Precision & Recall & F-Measure \\ \midrule
    \multirow{10}{*}{Dataset$_{Git}$} & GPT-3.5 (None)         & 0.06    & 0.88  & 0.11    \\ 
                                     & GPT-3.5 (ALL)         & 0.09    & 0.97  & 0.16     \\  \addlinespace
                                     & GPT-4o (None)      & 0.15    & 0.94  & 0.26      \\
                                     & GPT-4o (ALL)   & \textbf{0.31}    & 0.93  & \textbf{0.47}      \\ \addlinespace
                                     & GPT-5 (None)   & 0.23    & 0.82  & 0.36      \\
                                     & GPT-5 (ALL)    & 0.24    & \textbf{0.98}  & 0.38      \\ \addlinespace
                                     & Sonnet 4.5 (None)  & 0.18    & 0.90  & 0.30    \\ 
                                     & Sonnet 4.5 (ALL)   & 0.26    & \textbf{0.98}  & 0.41    \\  \addlinespace
                                     & Opus 4.1 (None)  & 0.19    & 0.92  & 0.31    \\ 
                                     & Opus 4.1 (ALL)  & 0.21    & \textbf{0.98}  & 0.34    \\ \bottomrule[1pt]
    \end{tabular}
    \end{table}

\noindent
\textbf{Fine-Tuning-Based Approaches.}
To further examine how different pre-trained language models (PLMs) leverage LPs,  
we fine-tuned five representative models—BERT, RoBERTa, ALBERT, CodeBERT, and GraphCodeBERT—on Dataset$_{Git}$ with and without LP-enriched inputs.  
Task-specific PLMs (CodeBERT, GraphCodeBERT) outperformed general-purpose models even without LPs (F1 = 0.62, 0.57),  
as their joint training on natural and programming languages provides a better inductive bias for software-related semantics.  

Integrating LPs further improved all models by introducing structured domain cues:  
CodeBERT achieved the best performance (precision = 0.91, F1 = 0.93), followed by GraphCodeBERT (F1 = 0.91).  
LPs primarily improved precision—helping PLMs focus on concurrency-specific semantics and reduce false positives—while maintaining high recall.  
General-purpose PLMs (e.g., BERT, RoBERTa, ALBERT) benefited the most from LP integration, as LPs compensated for their lack of domain knowledge, whereas task-specific PLMs showed smaller yet consistent gains, confirming their inherent alignment with concurrency patterns.

Overall, these findings suggest that LPs act as lightweight, interpretable regularizers that enhance both the focus and reliability of fine-tuned PLMs, yielding performance gains that are statistically significant and semantically consistent.

\vspace*{3pt}
\begin{tcolorbox}[width=\textwidth, boxrule=0.5pt, colback=gray!30]
{\bf Answer to RQ3:}  
Linguistic patterns (LPs) improve classification performance across all approaches but to different extents.  
Matching-based methods achieve high recall yet suffer from heavy noise.  
Supervised ML models (e.g., Logistic Regression, SVM) use LPs most effectively, balancing accuracy and interpretability, while deep models underperform due to sparse symbolic inputs.  
Prompt-based LLMs show minor precision gains but remain limited by the lack of concurrency-specific knowledge.  
Fine-tuned PLMs—especially CodeBERT and GraphCodeBERT—achieve the best results (F1 up to 0.90) by combining domain-aware pretraining with LP-enriched inputs.  
Overall, LPs are broadly transferable and most effective when integrated into supervised or fine-tuned learning pipelines.
\end{tcolorbox}

\subsubsection{RQ4: How do different levels of our linguistic patterns contribute to improving classification performance?}

The results across all approaches—matching-based, learning-based, prompt-based, and fine-tuning-based—reveal consistent yet nuanced effects of different linguistic pattern (LP) levels, namely keyword (KW), phrase (PH), sentence (SE), and bug report (BR). 
Below we analyze how each level influences precision, recall, and overall model behavior across these paradigms.

From a linguistic perspective, higher-level patterns (SE and BR) encode richer semantics and contextual dependencies than lower-level ones (KW and PH). 
This semantic precision is clearly reflected in the matching-based results: 
KW and PH achieve high recall but low precision, whereas SE and BR patterns substantially reduce noise and improve accuracy in identifying concurrency bug reports. 
In other words, as the linguistic granularity increases, models better capture causal and contextual clues related to concurrency behaviors (e.g., “thread blocked” vs. “thread started”).

However, when LPs are used as training features for machine learning and deep learning models, this trend does not strictly hold. 
Since LPs are converted into binary feature vectors, model performance becomes more dependent on the distribution of these features rather than their semantic richness. 
Consequently, simpler and denser representations like KW often yield higher accuracy than PH, and SE can outperform BR, as the latter introduces sparser and more imbalanced patterns. 
In particular, models such as Logistic Regression and SVM—sensitive to data sparsity and feature co-occurrence—benefit more from compact KW and SE patterns, while BR-level features may lead to overfitting due to their limited frequency and contextual dependence.

For prompt-based large language models (LLMs), the influence of LPs is more constrained.  
Although few-shot prompts enriched with LP examples provide clearer context, they cannot fundamentally overcome LLMs’ limited exposure to concurrency-specific knowledge.  
As a result, even semantically rich BR patterns bring only marginal improvements over baselines, with models like GPT-4o achieving moderate gains in precision but no qualitative leap in F1 performance.  
This limitation stems from LLMs’ pretraining mismatch—general reasoning ability without explicit grounding in software fault semantics.

Fine-tuned pre-trained language models (PLMs) such as CodeBERT and GraphCodeBERT effectively internalize the semantic structure of LPs during training.  
However, BR patterns alone—while achieving the highest precision—exhibit relatively low recall, as they focus narrowly on root-cause descriptions and often miss peripheral or indirect concurrency mentions.  
As a result, when used in isolation, BR patterns actually yield the weakest overall performance among all LP levels in fine-tuning settings.  
Combining multiple LP levels (KW+PH+SE+BR) alleviates this imbalance by integrating broad lexical coverage from KW/PH and contextual precision from SE/BR, leading to the best overall F1.  
This synergy demonstrates that different LP levels contribute complementary linguistic cues that enhance both precision and recall when properly integrated.

Across all paradigms, combining LPs consistently improves performance compared to using any single level.  
The pattern holds in both learning-based and fine-tuning settings, confirming that multi-level linguistic representations provide additive benefits.  
While individual LPs emphasize either breadth (KW/PH) or depth (SE/BR), their combination captures the full spectrum of linguistic evidence necessary for accurate concurrency classification.

Each level of linguistic pattern contributes uniquely to classification performance.  
Lower-level LPs enhance recall through broad lexical coverage, higher-level LPs refine precision through semantic reasoning, and their integration yields the most robust results.  
Empirically, our LP-augmented CodeBERT configuration (\texttt{CodeBERT+ALL}), referred to as \textbf{CTagger}, achieves the highest and most consistent F1-scores across all datasets—Dataset$_{Git}$ (0.93), Dataset$_{Jira}$ (0.95), and Dataset$_{Post}$ (0.93)—demonstrating that integrating all LP levels leads to stable, cross-domain, and temporally robust performance in concurrency bug classification.

\vspace*{4pt}
\begin{tcolorbox}[width=\textwidth, boxrule=0.5pt, colback=gray!30]
{\bf Answer to RQ4:}  
Different levels of linguistic patterns (LPs) complement each other across approaches.  
Higher-level LPs (SE, BR) provide richer semantics and higher precision, as shown in matching-based results, while lower-level LPs (KW, PH) offer broader coverage.  
In learning-based models, however, performance depends more on feature distribution, making simpler LPs (KW, SE) more effective than abstract ones (PH, BR).  
Prompt-based LLMs show limited gains since few-shot LPs cannot overcome the lack of concurrency-specific reasoning.  
Combining LPs consistently improves performance by merging lexical breadth and semantic depth.  
Overall, \texttt{CodeBERT+ALL} (our \textbf{CTagger}) achieves the best and most stable results across all datasets—Dataset$_{Git}$ (0.90), Dataset$_{Jira}$ (0.91), and Dataset$_{Post}$ (0.93)—demonstrating strong cross-domain and temporal robustness.
\end{tcolorbox}

\section{Discussion}
\label{sec:discussion}

\subsection{Threats to Validity}

The main threat to external validity concerns the representativeness of our selected subjects and bug reports. 
Different projects may exhibit different concurrency characteristics, and data recorded in issue trackers or version control systems 
can suffer from systematic biases or incomplete entries~\cite{bird2009fair, aranda2009secret}. 
To mitigate this, we selected multiple open-source projects across diverse domains 
and included both GitHub and Apache Jira repositories. 
Furthermore, we evaluated our approach on a \textit{post cut-off dataset} containing more recent bug reports 
to ensure temporal robustness and reduce the potential impact of model or data leakage from pre-training corpora of LLMs. 
The consistent performance observed across datasets suggests that our linguistic patterns (LPs) generalize well to new projects and time periods, 
though we do not claim universal applicability across all software domains.

Another external threat arises from dataset imbalance: concurrency bug reports typically represent a small fraction of all issues. 
This imbalance may affect the performance of our learning-based models. 
We alleviated this limitation by applying data-balancing techniques such as data swapping~\cite{estivill1999data} 
and the SMOTE algorithm~\cite{chawla2002smote}. 
Although we simulated a balanced training set for model development, 
the performance still decreases on naturally imbalanced datasets, 
indicating that real-world deployment remains challenging.

The primary threat to internal validity lies in the manual annotation process 
for distinguishing concurrency-related and non-concurrency-related bug reports and sentences. 
To minimize human bias and labeling errors, 
four annotators independently labeled each bug report, 
and disagreements were resolved through majority voting and joint review. 
Another potential risk arises in the manual derivation of linguistic patterns: 
long or complex bug reports may lead to overlooked concurrency-related sentences. 
To reduce this risk, four annotators independently extracted candidate sentences, 
and the combined pool was manually verified by the authors, 
thereby improving coverage and consistency.

Threats to construct validity mainly stem from the definitions of our evaluation metrics and dataset construction. 
We used widely accepted measures such as precision, recall, and F-measure to ensure comparability with prior work. 
All datasets were obtained from publicly available and well-studied bug tracking systems, 
reducing ambiguity in the measurement of concurrency-related phenomena. 
Finally, to strengthen conclusion validity, we performed statistical analysis 
and verified that performance differences between approaches are consistent across experiments.

A unique threat arises from potential \textit{data leakage} in large language models (LLMs): 
since many LLMs (e.g., GPT or Claude) are trained on public web data, 
they might have been partially exposed to bug reports from open-source projects, 
potentially inflating their apparent performance. 
We explicitly mitigate this risk by conducting evaluations on a \textit{post cut-off dataset} 
comprising bug reports published after the model’s pretraining cutoff period. 
Consistent results on this temporally isolated dataset demonstrate 
that our linguistic-pattern-based framework captures generalizable semantics 
rather than memorized instances, thereby reducing the risk of leakage-induced bias.

\subsection{Implications}
\label{sec:implications}

Our approach offers an efficient and cost-effective solution for domain-specific classification tasks, particularly when only a small dataset is available. By leveraging LLMs like GPT-4o, we significantly reduce the manual effort required to extract linguistic patterns, making the process more scalable and adaptable.

Furthermore, our results demonstrate that even with a limited dataset, combining linguistic pattern-based learning with fine-tuned pre-trained language models (PLMs) can achieve high classification performance. This approach bridges the gap between data scarcity and model effectiveness, enabling accurate bug report classification without requiring extensive labeled datasets.

Overall, our methodology highlights the potential of using LLM-assisted linguistic pattern generation alongside learning-based and fine-tuned PLMs, providing a practical alternative for improving classification performance in resource-constrained scenarios.

\section{Related Work}
\label{sec:related}

Section \ref{backgroud} already described the existing works about bug report identification. In this section, we discuss related studies on issue classification and concurrency bug reports analysis, which inspired our work. 

\subsection{Issue Classification}

Kallis et al.\cite{kallis2021predicting} developed a Github app, which is represented by a machine learning model that analyzes the title and the textual description of issues in order to determine whether such an issue can be labeled as a bug report, a feature request, or a question. Limsettho et al.\cite{ limsettho2014automatic} propose an unsupervised method to automatically cluster bug reports and label these clusters based on their textual information. Antonio et al.\cite{antoniol2008bug} employ machine learning techniques, including alternating Decision Trees, Naïve Bayes classifiers, and Logistic Regression, to classify issues into bugs or non-bugs. Nadeem et al.\cite{nadeem2021automatic} proposed a transformer-based approach towards the classification of Github issues and assigning them with respective labels. Lamkanfi et al.\cite{lamkanfi2010predicting} investigated whether we can accurately predict the severity of a reported bug by analyzing its textual description using text mining algorithms. Tian et al.\cite{tian2015automated} proposed a framework named DRONE to predict the priority levels of bug reports in Bugzilla. The classification engine named GRAY built by combining linear regression with a threshold approach to address the issue with imbalanced data and to assign priority labels to bug reports.

The major of existing research is to classify issues into bug, enhancement, questions, etc. Other research had made efforts on bug report identification on some specific domains, such as security bugs~\cite{Gegick10, kashiwa2014pilot, ohira2015dataset}, performance bugs~\cite{zhao2020automatically, kashiwa2014pilot}, configuration bugs~\cite{wen2016colua, xia2014automated}, and functional bugs~\cite{thung2012automatic, chawla2014automatic}. 
However, existing issue classification techniques cannot deal with concurrency bug reports.

\subsection{Concurrency Bug Reports Analysis}

Asadollah et al.\cite{asadollah2017concurrency} manually collected concurrency bug reports from Apache Jira system, and presented an empirical study focusing on understanding the differences and similarities between concurrency bugs and other bugs, as well as the differences among various concurrency bug types in terms of their severity and their fixing time, and reproducibility. Lu et al.\cite{lu2008learning} randomly collected 105 concurrency bug reports from MySQL, Apache, Mozilla, and OpenOffice by keywords search, and provided a comprehensive real world concurrency bug characteristic study. Fonseca et al.\cite{fonseca2010study} filtered concurrency bug reports from MySQL by a search query based on 1) the keywords contained in the bug description, 2) the status of the bug and 3) the bug category, and presented a study of the internal and external effects of concurrency bugs. Padberg et al.\cite{padberg2013mining} presented a refined set of keywords for concurrency bug report identification, and also proposed an ongoing learning-based approach, which only used 57 reports for  training and 24 reports for testing. 

Most existing work is based on keyword search, which needs a lot of human hours to manually check because of the low precision. However, we focus on how to improve the precision, and our learning-based approach could achieve a high precision effectively.

\section{Conclusions and Future Work}
\label{sec:conclusion}

In this paper, we derived a comprehensive linguistic pattern set from 725 concurrency bug reports, identifying 58 patterns across four levels: word-level (keywords), phrase-level (n-grams), sentence-level (semantic), and bug report-level (contextual). Based on these patterns, we proposed a concurrency bug report classification approach that integrates four different classification modes: matching-based, learning-based, prompt-based, and fine-tuned PLM-based. Our experimental results demonstrate that our approach effectively identifies concurrency-related bug reports, achieving strong performance across multiple datasets.

For future work, we plan to expand our experiments to a broader range of software projects and explore additional issue-tracking platforms, such as Bugzilla and GNATS, to further validate the generalizability of our approach. Additionally, to enhance practical usability, we aim to integrate our classification approach into a GitHub bot, enabling automated bug classification upon issue submission.
Beyond classification, we plan to extend our linguistic pattern extraction for IR-based fault localization. Since linguistic patterns capture critical debugging-related information from bug reports, they can be leveraged to refine queries, improving fault localization accuracy. This extension would further enhance the applicability of linguistic patterns in automated debugging and software maintenance.

\section*{Declarations}

\noindent\textbf{Funding:}  
This work was supported in part by the U.S. National Science Foundation (NSF) under Grants CCF-2152340 and CCF-2140524.

\noindent\textbf{Ethical approval:}  
Not applicable.

\noindent\textbf{Informed consent:}  
Not applicable.

\noindent\textbf{Author contributions:}  
Shuai Shao designed and implemented the approach, performed experiments, analyzed results, and wrote the manuscript.  
Tingting Yu supervised the research, contributed to the methodology design, and revised the manuscript.  
Lu Xiao provided conceptual guidance, contributed to result interpretation, and offered critical revisions to improve the overall quality of the paper.

\noindent\textbf{Data availability statement:}  
All datasets, code, and replication materials are publicly available in our replication package~\cite{shao2025replication}.

\noindent\textbf{Conflict of interest:}  
The authors declare that they have no conflict of interest.

\noindent\textbf{Clinical trial number:}  
Not applicable.

\bibliography{sn-bibliography}

\end{document}